\documentstyle[preprint,prb,aps]{revtex}
\tighten
\begin{document}
\title{Spin, charge and orbital ordering in ferrimagnetic insulator YBaMn$_2$O$_5$}
\author{R.~Vidya$^*$, P.~Ravindran, A.~Kjekshus and H.~Fjellv{\aa}g}
\address{ Department of Chemistry, University of Oslo, Box 1033, Blindern,
N-0315 Oslo, Norway.}
\date{\today}
\maketitle
\begin{abstract}
{ The oxygen-deficient (double) perovskite YBaMn$_2$O$_5$, containing corner-linked MnO$_5$
square pyramids, is found to exhibit ferrimagnetic ordering in its ground state.
In the present work we report generalized-gradient-corrected, relativistic
first-principles full-potential density-functional calculations performed on 
YBaMn$_2$O$_5$ in the nonmagnetic, ferromagnetic and ferrimagnetic states. 
The charge, orbital and spin orderings are explained with site-, angular
momentum- and orbital-projected density of states, charge-density plots, electronic
structure and total energy studies. 
YBaMn$_2$O$_5$ is found to stabilize in a G-type ferrimagnetic
state in accordance with experimental results. The experimentally observed insulating
behavior appears only when we include ferrimagnetic ordering in
our calculation. We observed significant optical anisotropy in this material
originating from the combined effect of ferrimagnetic ordering and
crystal field splitting. In order to gain knowledge about the presence
of different valence states for Mn in YBaMn$_2$O$_5$ we have calculated  $K$-edge
x-ray absorption near-edge spectra for the Mn and O atoms.
The presence of the different valence states for Mn is clearly established from the 
x-ray absorption near-edge spectra, hyperfine field parameters and the magnetic 
properties study. Among the experimentally proposed structures, the recently 
reported description based on $P$4/$nmm$ is found to represent the stable structure.
}
\end{abstract}
\section{INTRODUCTION}
\label{sec:intro}
Perovskite-type transition-metal oxides {\em ABO$_3$} and their oxygen-deficient
relatives exhibit a variety of interesting physical properties including high-temperature
superconductivity, metal-insulator transition and a variety of cooperative magnetic 
phenomena. Among them
manganites have recently attracted particular attention because of the discovery of colossal
negative magnetoresistance (CMR) in La$_{1-x}$Sr$_x$MnO$_3$,\cite{fujimori98}
La$_{1-x}$Ba$_x$MnO$_3$\cite{fujimori98} and related 
phases.\cite{mahendiran95,liu95,gong95,caignart95} This renewed interest 
in the mixed-valence manganese perovskites such as La$_{1-x}$Pb$_x$MnO$_3$ is due 
to their potential technological applications.\cite{furukawa94,millis95} 
In addition, the search for new high-temperature superconductors in
mixed-oxide materials is a driving force for attention. The mechanism of high-temperature 
superconductivity is believed to be
linked to cooperative interaction between copper 3$d_{x^2-y^2}$ and 3$d_{z^2}$
orbitals and oxygen $2p$ orbitals.\cite{dixon91} An attractive approach to obtain insight
in the nature of this phenomenon is to examine the magnetic and electrical properties of
non-copper oxide analogues of known high-temperature superconducting cuprates. 
Manganese is a good choice for such a task, as in an octahedral perovskite-like 
configuration, Mn$^{3+}$ ($d^{4}$ high spin with a single electron in an $e_g$  orbital)
will experience a similar Jahn-Teller (JT) distortion to that of Cu$^{2+}$ 
[$d^9$ high spin with three electrons (or one hole) in the $e_g$ orbitals]. 
The hole-doped manganese perovskites show some similarities 
to the corresponding hole-doped Cu phases in which superconductivity 
occurs.\cite{torrance89} Structural similarities between these two groups of materials 
suggest that new Mn analogues of the high-temperature superconducting Cu oxides 
may be prepared.
\par
Since the discovery of CMR phenomena in
perovskite-related manganites, extensive studies have been performed on manganese oxides
with atomic arrangements related to the perovskite and pyrochlore stuctures
over a wide variety of compositions with the aim to explore exotic
spin-charge coupled state.\cite{kuwahara98} Anisotropic CMR phenomena have also recently 
been reported in layered Ruddlesden-Popper variants of
perovskites ($RE,AE)_{n+1}$Mn$_n$O$_{3n+1}$ ($RE$ = rare earth;
$AE$ = alkaline earth)\cite{moritomo96} for $n$ = 2. Similar effects have also been observed in the oxygen-deficient cubic pyrochlore Tl$_2$Mn$_2$O$_{7-\delta}$. The chemical features
common to these materials are an intimately connected Mn-O-Mn network, within a
three-dimensional or multi-layered structure, and an average Mn oxidation state between +3
and +4 (obtained by hole-doping of Mn$^{3+}$).\cite{chapman96}
\par 
At low temperatures, manganese perovskites are characterized by strong
competition between charge-carrier itineracy and localization. In the former case
a ferromagnetic (F) metallic state is formed. In the latter case, the localized 
carriers tend to form charge-ordered (CO) states, which have a predominantly 
antiferromagnetic (AF) insulating character.
Hence, we have a competition between F with metallic behavior and cooperative
JT distortion with CO. The CO state can be converted into a
F metallic state, by the application of a magnetic field.
The intriguing doping-induced, temperature-dependent metal-insulator transition
and the interwoven magnetic (spin), orbital and charge-ordering phenomena in mixed-valence 
manganese perovskites and transition-metal oxides have 
attracted much attention in recent years. \cite{vogt2000} An active role of 
the orbital degree of freedom in the lattice and electronic response can be most typically
seen in manganese perovskite oxides.
As a matter of fact such properties appear to have their origin in the
unique electronic structures derived from the hybridized Mn $3d$ and O $2p$ orbitals
in the particular structural and chemical environment of a perovskite. The thus resulting 
intra-atomic exchange and the orbital degrees of freedom of the Mn $3d$ electrons 
play essential roles in this constellation. Furthermore,
various kinds of structural distortion profoundly influence the electronic properties.
The extensive study of CMR in $RE_{1-x}$$AE_x$MnO$_3$ have brought forth novel 
features related to CO in these oxides. 
In transition-metal oxides with their anisotropic-shaped $d$ orbitals, 
Coulomb interaction between the electrons (strong electron correlation effect)
may be of great importance. Orbital ordering (OO) gives rise to 
anisotropic interactions in the electron-transfer process which in turn favors or 
disfavors the double-exchange and the superexchange
(F or AF) interactions in an orbital direction-dependent manner and 
hence gives a complex spin-orbital-coupled state. OO in the manganese 
oxides occasionally accompanies the concomitant CO.\cite{toku_sci}
The ordered oxygen-deficient double perovskites $RE$Ba$T$$_2$O$_{5+\delta}$ 
($T$ = transition metals like Fe, Co and Mn) have attracted much attention as 
new spin-charge-orbital coupled systems and new CMR materials. 
In the isostructural phases with $RE$ = Gd and Eu CMR effects of some 40\% 
are observed.\cite{martin97}
As the experimental findings have been made available only in recent years,
little theoretical work has been undertaken to understand the origin of these
microscopic properties. The present study reports a detailed theoretical investigation 
on the electronic structure and optical properties of YBaMn$_2$O$_5$.
\par
At low temperature, YBaMn$_2$O$_5$ is an AF insulator with  CO of Mn$^{3+}$ and 
Mn$^{2+}$ accompanied by OO and spin ordering (SO).\cite{millange99} The mechanism 
of CO and SO in manganites is not at all clear. Different authors have emphasized the 
importance of different ingredients such as on-site Coulomb 
interactions,\cite{anisimov97,koshibae97,mizokawa97} JT distortion\cite{lee98,solovyev96} 
and inter-site Coulomb interactions.\cite{lee97} Therefore, we have attempted to study CO, OO and SO 
through full potential linear muffin-tin orbital (FPLMTO) and full potential
linear augmented plane wave (FPLAPW) methods.
Similar to Fe$_3$O$_4$ and SmBaFe$_2$O$_{5+w}$ where Fe takes the conventional
valence states of Fe$^{3+}$ and Fe$^{2+}$ at low temperature,\cite{cox92,linden99}  
Mn is reported to occur as Mn$^{3+}$ and Mn$^{2+}$ in YBaMn$_2$O$_5$. Above the
so called Verwey temperature ($T_V$) valence-state mixing has been observed in 
Fe$_3$O$_4$ as well as in SmBaFe$_2$O$_{5+w}$. This brings in an additional 
interesting aspect in the study of the electronic
structure and magnetic properties of Mn in YBaMn$_2$O$_5$. The
first structure determinations based on powder x-ray (300\,K)\cite{chapman96} and 
neutron (100$-$300\,K)\cite{mcallister98} diffraction data found that YBaMn$_2$O$_5$ 
crystallizes in space group 
$P$4/$mmm$, whereas a more recent powder neutron (2$-$300\,K)\cite{millange99} 
diffraction study (PND) found $P$4/$nmm$. So, a theoretical examination of the total energy for 
the two different structural alternatives is required. 
\par 
The rest of the paper is arranged as follows: Sec.\,\ref{sec:str} gives crystal structure 
details for YBaMn$_2$O$_5$. The theoretical methods used for the 
calculations are described in Sec.\,\ref{sec:comp}. The analysis of the band structure 
is given in Sec.\,\ref{subsec:band}. Sec.\,\ref{subsec:dos}  deals 
with the nature of the chemical bonding in YBaMn$_2$O$_5$, analyzed with 
the help of site-, angular momentum- and orbital-projected density of states (DOS). 
Sec.\,\ref{subsec:charge} discusses CO and OO  
in YBaMn$_2$O$_5$ with the help of charge density plots. The results from 
calculations of optical spectra and x-ray absorption near edge (XANE) spectra 
are discussed in Sec.\,\ref{subsec:optic} and \ref{subsec:xane} respectively. 
Sec.\,\ref{subsec:hyp} deals with hyperfine parameters. Finally the 
important conclusions are summarized in Sec.\,\ref{sec:sum}.

\section{CRYSTAL STRUCTURE}
\label{sec:str}
Chapman {\em et al.}\cite{chapman96} synthesized rather impure YBaMn$_2$O$_5$ and 
reported the crystal structure parameters
according to space group $P$4/$mmm$ [described as double or ordered, oxygen-deficient 
perovskite; closely related to YBaCuFeO$_5$]. These findings were subsequently confirmed by
McAllister and Attfield\cite{mcallister98} who also establised a model for the ferrimagnetic 
ordering  of the magnetic moments of Mn in YBaMn$_2$O$_5$ (still based on an impure sample).
More recently Millange {\em et al.}\cite{millange99} have succeeded in preparing phase-pure
YBaMn$_2$O$_5$ and these authors report crystal and magnetic structure 
parameters according to space group $P$4/$nmm$.
\par
Within the $P$4/$nmm$ description YBaMn$_2$O$_5$ crystallizes  
with $a$ = 5.5359 and $c$ = 7.6151\,\AA; Y in 2(b), Ba in 2(a), Mn(1) in 2(c) with $z$ = 
0.2745, Mn(2) in 2(c) with $z$ = 0.7457, O(1) in 8(j) with $x$ = 0.4911, $y$ = 0.9911 and
$z$ = 0.4911 and O(2) in 2(c) with $z$ = 0.0061. The lacking oxygens in the
yttrium plane, compared with the perovskite-aristotype structure reduces the coordinate 
number of yttrium to 8, while barium retains the
typical twelve coordination of the perovskite structure. The Mn-O network consists of double
layers of MnO$_5$ square pyramids, corner shared in the $ab$ plane and linked {\it via} their 
apices. 
\par 
According to the $P$4/$nmm$ description (Fig.\,\ref{fig:struc}) the two kinds of
MnO$_5$ pyramids are arranged in an ordered manner, each Mn$^{2+}$O$_5$ pyramid 
being linked to five Mn$^{3+}$O$_5$ pyramids. Owing to this charge ordering, 
each Mn$^{2+}$ has four Mn$^{3+}$ in-plane neighbors.
Oxygen takes two crystallographically different sites. O(1)  
forms the base of the square pyramids, while O(2) is located at the apex of the pyramids. 
The interatomic Mn-O distances fall in four categories, being 1.908 
and 2.086 {\AA} for Mn$^{3+}$-O(1) and Mn$^{2+}$-O(1), respectively, whereas 
O(2) is 2.081 and 1.961 {\AA} away from Mn$^{3+}$ and Mn$^{2+}$, respectively.
(The $P$4/$mmm$ description gave on the other hand, almost equal
basal and apical Mn-O distances within all square-pyramids.\cite{chapman96})
The basal plane Mn$^{3+}$-O(1)-Mn$^{2+}$ angle is 157.8$^o$ and the apical 
Mn$^{3+}$-O(2)-Mn$^{2+}$ angle is 180$^o$. The large variation in these angles 
play a key role in the magnetic properties.

\section{Computational Details}
\label{sec:comp}
\subsection{The FPLMTO computations}
The present calculations have used the full-potential linear muffin-tin orbital 
(FPLMTO) method\cite{wills} where no shape approximation is assumed for the 
one-electron potential and charge density. The basis geometry consists of muffin-tin (MT)
spheres centered around the 
atomic sites with an interstitial region in between. Inside the MT spheres the charge density 
and potential are expanded by means of spherical harmonic functions multiplied by
a radial component. The interstitial basis function is a Bloch sum of linear combinations
of Neumann or Henkel functions depending on the sign of the kinetic energy $\kappa^{2}$
(corresponding to the basis functions in the interstitial region). 
Each Neumann or Henkel function is then augmented (replaced) by a numerical basis function 
inside the MT spheres, in the standard way of the linear MT orbital method.\cite{andersen75} 
Since a Bloch sum of atomic centered Henkel or Neumann functions  
has the periodicity of the underlying lattice it may be expanded in a Fourier 
series, as done here. The spherical-harmonic expansion of the
charge density, potential and basis functions was performed up to $\ell_{max}$ = 6. 
The basis included Y 
$4p$, $5s$, $5p$ and $4d$ states, Ba $5p$, $6s$, $6p$, $5d$ and $4f$ states, 
Mn $4s$, $4p$ and $3d$ states and O $2s$, $2p$ and $3d$ states.
\par
Furthermore, the calculations are all-electron as well as fully relativistic. The latter 
level is obtained by including the mass velocity and Darwin (and higher order) terms in the
derivation of the radial functions (inside the MT spheres) whereas the spin-orbit 
coupling was included at the variational step using an ($\ell$,$s$) basis. Moreover, 
the present 
calculations made use of a so-called double basis, to ensure a well-converged wave 
function. 
This means that two Neumann or Henkel functions were applied, each attached to its 
own radial
function with an ({\em n},$\ell$) quantum number. The integrations over the 
Brillouin zone (BZ) in the ground state calculations were performed as a weighted sum, 
using the special point sampling,\cite{chadi89} with weights reflecting the symmetry of a 
given $\bf{k}$ point. We also used a Gaussian smearing width of 20\,mRy for each eigenvalue 
in the vicinity of the
Fermi level to speed up the convergence. For the DOS and optical calculations, the tetrahedron
integration was employed. The calculations were performed for the experimentally determined 
structural parameters (see Sec.\,\ref{sec:str}). 
For the exchange-correlation functional, $E_{xc}(n)$, we have used the generalized 
gradient approximation (GGA) where the gradient of the electron density is taken into
account using Perdew and Wang\cite{perdew96} implementation of GGA.
192 {\bf k} points in the irreducible part of the
primitive tetragonal BZ were used for the self-consistent ground state 
calculations and 352 {\bf k} points for the optical calculations.

\subsection{The FPLAPW computations}
\label{subsec:flap}
For the XANES and orbital-projected DOS calculations we have
applied the full-potential linearized-augmented plane wave (FPLAPW) method\cite{blaha97}
in a scalar-relativistic version without spin-orbit coupling. The FPLAPW method
divides space into an interstitial region (IR) and non-overlapping MT
spheres centered at the atomic sites. In IR, the basis set consists of plane
waves. Inside the MT spheres, the basis set is described by radial solutions of the
one-particle Schr\"{o}dinger equation (at fixed energies), and their energy derivatives
multiplied by spherical harmonics. The charge densities and potentials in the atomic
spheres were represented by spherical harmonics up to $\ell$ = 6, whereas in the interstitial
region these quantities were expanded in a Fourier series with 3334 stars of the reciprocal
lattice vectors {\bf G}. The radial basis functions of each LAPW were calculated up to
$\ell$ = 10 and the non-spherical potential contribution to the Hamiltonian matrix had an
upper limit of $\ell$ = 4. Atomic-sphere radii $R_{MT}$ of 2.5, 2.8, 1.8 and 1.6\, a.u. 
for Y, Ba, Mn and O, respectively, were used. Since the spin densities are well confined 
within a radius of about 1.5 a.u, the resulting magnetic moments do not depend appreciably
with the chosen atomic-sphere radii. The initial basis set included 5$s$, 5$p$, 4$d$ 
valence and 4$s$, 4$p$ semicore functions for Y, 6$s$, 6$p$, 6$d$ valence and 5$s$, 
5$p$ semicore
functions for Ba, 4$s$, 4$p$, 3$d$ valence and 3$s$, 3$p$ semicore functions for 
Mn and 2$s$, 2$p$ and 3$d$ functions for O. These basis functions 
were supplemented with local orbitals\cite{dsingh} for additional flexibility to the 
representation of the semicore states and for generalization of relaxation of the 
linearization errors.
Owing to the linearization errors DOS are reliable only to about 1 to 2\,Ry above $E_{F}$.
Therefore, after selfconsistency was achieved for this
basis set we included higher energy local orbitals: $5d$- and $4f$-like function for Y,
$6d$- and $4f$-like function for Ba, $5s$- and $5p$-like functions for Mn and $3p$-like 
functions for O. The BZ integration was done with a modified tetrahedron
method\cite{blochl94} and we used 140 {\bf k} points in the irreducible wedge of BZ.
Exchange and correlation effects are treated within density-functional
theory (DFT), using GGA.\cite{perdew96}

\subsection{Optical properties}
\label{subsec:opt}
Optical properties of matter can be described by means of the transverse dielectric
function {\bf $\epsilon$( q}, $\omega$) where {\bf q } is the momentum transfer in the 
photon-electron interaction and $\omega$ is the energy transfer. At lower energies one can set 
{\bf q} = {\bf 0}, and arrive at the electric dipole approximation, which is assumed 
throughout this paper. The real and imaginary parts of {\bf $\epsilon$}($\omega$) are 
often referred to as {\bf $\epsilon$$_1$} and {\bf $\epsilon$$_2$}, respectively. 
We have calculated the dielectric function for frequencies well above those of 
the phonons and therefore we considered only electronic excitations.  
In condensed matter systems, there are two contributions to
{\bf $\epsilon$}($\omega$), {\em viz.} intra-band and inter-band transitions. The 
contribution from intra-band transitions is important only for metals. The inter-band 
transitions can further be split into direct and indirect transitions. The latter 
involves scattering of phonons and are neglected here, and moreover these only make 
small contribution to {\bf $\epsilon$}($\omega$) in comparison to the direct 
transitions,\cite{smith71} but have a temperature broadening effect.
Also other effects, {\em e.g.,} excitons (which normally give rise to rather sharp peaks) 
affect the optical properties.
\par
The direct inter-band contribution to the imaginary part of the dielectric function,
{\bf $\epsilon$$_2$}($\omega$) is calculated by summing all possible transitions 
from occupied to unoccupied states, taking the appropriate transition-matrix 
element into account. The dielectric function is a tensor for which all components are
needed for a complete description. However, we restrict our considerations to the 
diagonal matrix elements $\epsilon$$^{{\nu}{\nu}}(\omega)$ with $\nu$ = $x$, $y$ or $z$. 
The inter-band contribution to
the diagonal elements of {\bf $\epsilon$}$_2$($\omega$) is given by
\begin{eqnarray}
\epsilon_{2}^{\nu\nu} = \frac{8\pi^{2}e^{2}}{m^{2}\omega^{2}} & & \times
\sum_{n}^{unocc} \sum_{n'}^{occ} \int_{BZ}|P_{nn'}^{\nu}(k)|^{2}
f_{kn}(1-f_{kn'})\delta(E_{n}^{k}-E_{n'}^{k}-\hbar\omega)\times
 \frac{d^{3}k}{(2\pi)^{3}},
\label{e2}
\end{eqnarray}
where $e$ is the electron charge, $m$ its mass, $f_{kn}$ the Fermi-Dirac distribution
function, $P_{nn'}^{\nu}$ the projection of the momentum matrix elements
along the direction $\nu$ of the electric field and $E_{n}^{k}$ one
electron energies. The evaluation of the matrix elements in Eq. (\ref{e2}) is done separately 
over the MT and interstitial regions. Further details about the
evaluation of matrix elements are found in Ref.\onlinecite{alouani96}. The integration 
over BZ in Eq. (\ref{e2}) is performed using linear interpolation
on a mesh of uniformly distributed points, {\em i.e.,} the tetrahedron method. The
total $\epsilon_{2}^{\nu\nu}$ was obtained from $\epsilon_{2}^{\nu\nu}(IBZ)$, 
{\em i.e.}
$\epsilon_{2}^{\nu\nu}$ was calculated only for the irreducible ($I$) part of BZ using
\begin{eqnarray}
{\bf \epsilon_{2}(\omega)} = \frac{1}{N} 
\sum_{i=1}^{N} {\bf \sigma_{i}}^{T} \epsilon_{2}(IBZ) {\bf \sigma_{i}},
\label{e22}
\end{eqnarray}
where $N$ is the number of symmetry operations and {\bf $\sigma_{i}$} represents
the symmetry operations; for shortness, $\epsilon(\omega)$ is used instead of 
$\epsilon^{\nu\nu}(\omega)$.
Lifetime broadening was simulated by convoluting the absorptive part of the dielectric
function with a Lorentzian, whose full width at half maximum (FWHM) is equal to  
$0.005(\hbar\omega)^2$\,eV. The experimental resolution was simulated by broadening the
final spectra with a Gaussian of constant FWHM equal to 0.01 eV.
\par
After having evaluated Eq. (\ref{e22}) we calculated the inter-band contribution to the
real part of the dielectric function $\epsilon_{1}(\omega)$ from the Kramers-Kronig 
relation
\begin{eqnarray}
{\bf \epsilon_{1}}(\omega) = 1 + \frac{2}{\pi}P 
\int_{0}^{\infty}\frac{{\bf \epsilon_{2}}(\omega'
)\omega'd\omega'}{\omega'^{2}-\omega^{2}}.
\label{e1}
\end{eqnarray}
In order to calculate  ${\bf \epsilon_{1}}(\omega)$ one needs a good
representation of  ${\bf \epsilon_{2}}(\omega)$ up to high energies. In the present 
work we have calculated ${\bf \epsilon_{2}}(\omega)$ up to 41 eV above the $E_{F}$ 
level, which also was the truncation energy used in Eq. (\ref{e1}). 
\par
To compare our theoretical results with the experimental spectra we 
have calculated polarized reflectivity spectra using the following relation. 
The specular reflectivity can be obtained from the complex dielectric constant 
in Eq. (\ref{e2}) through the Fresnel's equation, 
\begin{eqnarray}
R^{\nu\nu}(\omega) = \left|\frac{\sqrt{\epsilon(\omega)}-1}
{\sqrt{\epsilon(\omega)}+1}\right|^{2}.
\label{reflect}
\end{eqnarray}
\par
We have also calculated the absorption coefficient $I(\omega)$ 
and the refractive index $n$ 
using the following expressions:
\begin{eqnarray}
I(\omega) = 2\omega \left \lgroup \frac{(\epsilon_{1}^{2}(\omega)
+\epsilon_{2}^{2}(\omega))^{1/2} -\epsilon_{1}(\omega)}{2} \right \rgroup 
^{1/2}
\label{abs}
\end{eqnarray}
and
\begin{eqnarray}
n = \left \lgroup \frac{\sqrt{\epsilon_{1}^{2}+\epsilon_{2}^{2}}+\epsilon_{1}}
{2} \right \rgroup ^{1/2} .
\end{eqnarray}

\section{RESULTS AND DISCUSSION}
\label{sec:results}
\subsection{Electronic band structure}
\label{subsec:band}
The FPLMTO calculations were performed on YBaMn$_2$O$_5$ for three different
magnetic configurations, {\em viz.}, paramagnetic (P), ferromagnetic (F) and 
antiferromagnetic (AF).
From Table\,\ref{table:table1} it can be seen that in the AF configuration,
the spins are not cancelled and hence this state is really ferrimagnetic (Ferri).  
Moreover, Table\,\ref{table:table1} shows that Ferri YBaMn$_2$O$_5$ has lower energy than 
the P and F configurations. The energy-band structure of Ferri YBaMn$_2$O$_5$ is 
shown in Fig.\,\ref{fig:band}a and \ref{fig:band}b for up- and down-spin
bands, respectively.  YBaMn$_2$O$_5$ 
is seen to be an indirect-band-gap semiconductor. A closer inspection of the energy-band 
structure shows that the band gap is between the top of the valence band (VB) at the
$\Gamma$ point and the bottom of the conduction band (CB) at the $Z$ point.
As the unit cell contains 18 atoms, the band
structure is quite complicated and  Fig.\,\ref{fig:band} therefore only depicts energy 
range of $-$7.5 to 7.5\,eV. 
\par
There is a
finite energy gap of 1.307\,eV between the top-most occupied VB and the
bottom-most unoccupied CB in the up-spin channel. For the purpose of 
more clarity, it is convenient to divide the occupied portion of the band structure in the 
up-spin channel into three energy regions: (i) Bands lying at and below $-$4\,eV.
(ii) Bands lying between $-$4 and $-$2\,eV . (iii) The top of VB, closer to
$E_{F}$, {\em viz.}, the range  $-$2 to 0\,eV.
Region (i) contains 17 bands with contributions from Y $4s$, $5s$, Mn $3s$, $3d$ 
and O $2p$ electrons. Region (ii) comprises bands which originate from
completely filled O(1) and O(2) $2p$ orbitals. Region (iii) includes 10 bands. 
Among them one finds delocalized (dispersed) bands originating from Y $5s$ and O $2p$ 
orbitals and somewhat localized bands attributed to Mn(1) $d_{z^2}$, $d_{x^2-y^2}$, 
$d_{xz}$ and $d_{yz}$ orbitals. The top-most occupied band contains electrons 
stemming from the Mn $d_{z^2}$ orbital. In the unoccupied portion of the band structure, 
a corresponding division leads to two energy regions: (1) The bottom-most CB from 0 to 
2\,eV and (2) the middle range of CB from 2 to 4\,eV. (Above 4\,eV the bands are highly 
dispersed and it is quite difficult to establish the origin of the bands.)
There are 9 bands in region (1) which have some Y $5s$, Ba $5d$,
Mn(1) $3d_{xy}$ and Mn(2) $3d$ characters. In region (2) the bands retain 
Y $6s$, $4d$, Ba $7s$, $5d$ and Mn $4s$, $4p$ and $3d$ characters.
\par
The energy band structure of the down-spin channel (Fig.\,\ref{fig:band}b) has 16 bands in
the region (i) up to $-$4\,eV which arise from the $s$ and $p$ electrons of the Y, Ba, 
Mn and O atoms. The mainly $s$ and $p$ electron character of the
bands, makes them appreciably dispersed. The second energy region (ii) contains 12 
bands, which have Y $5s$, Ba $5p$, Mn(2) $3d$ and O(1), O(2) $2p$ character. The third 
region (iii) closer to $E_F$ has 10 bands which are 
mainly arising from Mn(1), Mn(2) $3d$ and O(1), O(2) $2p$ orbitals. Unlike the up-spin 
channel the down-spin channel contains two bands at $E_F$ which arise from the 
originally half-filled $t_{2g}$ orbitals of Mn(1) and the half-filled $d_{xy}$ orbital 
of Mn(2). 
\par
A finite band gap of 1.046\,eV opens up between the highest occupied VB and 
bottom-most unoccupied CB.
The unoccupied portion of the down-spin channel is quite different from that of 
the up-spin channel. The lowest-lying unoccupied band has Mn(2) $4s$ electrons.
Between 0 and 2\,eV there are 8 bands which arise mainly from
Mn(1) $3d$ electrons as well as from Mn(1), Mn(2) $4s$ electrons and O(1), O(2) $2p$ electrons.
The dispersed bands present between 2 and 4\,eV have Y $5s$, $3d$, Ba $6s$,
$5d$ and Mn(1) $3d$ characters. 

\subsection {DOS characteristics}
\label{subsec:dos}
In order to theoretically verify which of the two ($P4/mmm$ or $P4/nmm$ based) structures
is energetically more stable, we performed first-principle calculations for both
variants. The calculated DOS value at $E_F$ for the P phase $P$4/$mmm$ variant
is 192.82\,states/(Ry f.u.) and for $P$4/$nmm$ 149\,states/(Ry f.u.) in the P state. 
Hence, a larger number of electrons are present at $E_{F}$ for the former variant, which
favours the relative structural stability of the latter. 
Moreover, our calculations show that the $P4/nmm$ variant 
is 860\,meV/f.u. lower in energy than the $P4/mmm$ variant. Therefore we conclude 
that YBaMn$_2$O$_5$ is more stable in space group $P4/nmm$ than in
$P4/mmm$, {\em viz.} in accordance with the most recent PND-based experimental study.
\par
Our calculated total DOS curves for YBaMn$_2$O$_5$ in the P, F
and Ferri configurations are given in Fig.\,\ref{fig:totdos}. The highest 
occupied energy level in VB, {\em i.e.,} $E_{F}$ is marked by the dotted line. 
In the P and F cases finite DOS values are present in the vicinity of $E_{F}$. 
Hence, both the P and F configurations exhibit metallic character.
On going from the P to F case, the electrons start to localize which is seen 
from the reduced number of states at $E_{F}$. Due to the electron localization, the
gain in total energy of 3.1\,eV (Table\,\ref{table:table1}) is observed when the spin-polarization
is include in our calculations. However, in the Ferri 
case, a finite band gap of 0.88\,eV opens up between the top of VB and 
the bottom of CB. The experimental studies also show semiconducting 
behavior.\cite{chapman96,beales97} 
From Table\,\ref{table:table1} it can be seen that the Ferri 
configuration has lower energy than the other two configurations. The 
present observation of the stabilization
of a Ferri ground state in YBaMn$_2$O$_5$ is consistent with the established magnetic
structure.\cite{millange99} It is interesting to note that the introduction of the
Ferri configuration is essential in order to obtain the correct semiconducting ground 
state for YBaMn$_2$O$_5$. Unlike LaMnO$_3$ (where the energy difference between
the F and AF cases is $\sim$\,25\,meV)\cite{ravilamn} there is large energy difference
($\sim$\,0.5\,eV) between the Ferri and the F states of YBaMn$_2$O$_5$. So, a very large 
magnetic field is required to stabilize the F phase and induce insulator-to-metal 
transition in YBaMn$_2$O$_5$.
\par
In the $RE$MnO$_3$ phases hole doping induces CE-type magnetic ordering
in which the spins are F aligned in zigzag chains with AF coupling
between these chains. In YBaMn$_2$O$_5$, the Mn spins are AF aligned 
within quasi-one-dimensional chains as well as between the chains. The main difference
between YBaMn$_2$O$_5$ and $RE$MnO$_3$ is that the latter have $e_{g}$ electrons
present in the vicinity of $E_{F}$ and that the superexchange interaction originates
from the localized $t_{2g}$ electrons. Owing to the
square-pyramidal crystal field in YBaMn$_2$O$_5$ the $e_{g}$ electrons also get localized 
and hence both $e_{g}$ and the $t_{2g}$ electrons participate in the superexchange
interaction. This is the main reason why the F state has much higher energy than
the AF state in YBaMn$_2$O$_5$ as compared with LaMnO$_3$.
\par
To obtain more insight into the DOS features, we show the angular momentum- and site-decomposed 
DOS in Fig.\,\ref{fig:sitedos}. The lower panels for Y and Ba show that, 
in spite of the high atomic numbers for Y and Ba, small DOS values are seen in VB. The Y and Ba 
states come high up in CB ({\em ca.}\,4\,eV above $E_F$) indicating a nearly total 
ionization of these atoms. They lose their valence charge to form ionic 
bonding with oxygen. According to the crystal structure, Y and Ba are 
located in layers along $c$, which is clearly reflected in the electronic charge-density  
distribution within (110) in the AF configuration (Fig.\,\ref{fig:charge}). 
\par
The distinction between Mn(1) and Mn(2) is clearly reflected in 
the different topology of their DOS curves. As seen from Table\,\ref{table:table1}, 
Mn(1) has a smaller magnetic moment (3.07 \,$\mu_{B}$) than Mn(2) (3.91\,$\mu_{B}$)  
which leads to the conclusion that Mn(1) corresponds to Mn$^{3+}$ and Mn(2) to Mn$^{2+}$,
both in high-spin states. Both Mn(1) and Mn(2) have low-lying DOS features around 
$-$17\,eV
which can be attributed to well-localized $s$ and $p$ electrons originating from the 
covalent interaction between Mn $3d$ with O $2s$ and $2p$ states. The somewhat dispersed 
DOS in the energy region $-$7.5 to $-$2.5\,eV is attributed to
$3d$ electrons for both Mn(1) and Mn(2). As Mn(1) and Mn(2) are AF
coupled, their $d$-orbital DOS have opposite character. The localized peaks
closer to $E_{F}$, with a width of about 2\,eV,
have both $(3d)$ $e_{g}$ and $t_{2g}$ character (see below). 
\par
The top two panels of Fig.\,\ref{fig:sitedos} show DOS for O(1) and O(2). 
Crystallographically O(1) is co-planar (in $ab$) with Mn, whereas and O(2) is at the 
apex of the square pyramid (along $c$). The well-localized peaks in DOS for O(1) 
and O(2) at about $-$18\,eV reflect the completely filled $2s$ orbitals. The spread-out 
DOS features between
$-$7.5 to $-$0.5\,eV are attributed to O $2p$ states. Fig.\,\ref{fig:sitedos}\
shows that the O $2p$ states are energetically degenerate with Mn $3d$
states in this energy range, implying that these orbitals form covalent bonds with Mn(1)
and Mn(2) through hybridization. The almost empty DOS for O(1)
and O(2) in CB implies that the oxygen atoms are in nearly completely ionized states 
in YBaMn$_2$O$_5$. 
\par
In order to progress further in the understanding of the chemical bonding, charge, spin 
and orbital ordering in YBaMn$_2$O$_5$, we have plotted the orbital-decomposed DOS for 
the $3d$-orbitals
of Mn(1) and Mn(2) in Fig.\,\ref{fig:orb-dos}. This illustration shows that DOS for 
the $d_{z^2}$ 
orbital for both Mn(1) and Mn(2) are well-localized. There is a sharp peak at 
$-$5\,eV in the up-spin panel for Mn(1) and in the down-spin panel for Mn(2) which 
correlates with a well-localized peak in DOS for O(2) (Fig.\,\ref{fig:sitedos}) 
This is attributed to 
the 180$^o$ Mn$^{3+}$-O(2)-Mn$^{2+}$ bond angle which facilitates  $p$-$d$ $\sigma$ 
bond to the O(2) $p_{z}$ orbital and superexchange interaction.\cite{goodenough69} 
As up-spin Mn(1) and 
down-spin Mn(2) are involved, we infer that the superexchange interaction results 
in AF spin ordering between the Mn atoms involved. The peaks at {\em ca.} 1\,eV 
in up-spin Mn(1) and down-spin Mn(2) are attributed to the (non-bonding) $d_{z^2}$ orbitals.
\par
Turning to the other $e_{g}$ orbitals (of $d_{x^2-y^2}$ character) for Mn(1) and Mn(2), 
these are
spread in the $ab$ plane. From Fig.\,\ref{fig:sitedos}, we see that the O $2p$
orbitals are also situated in the same energy range ($-$5\,eV to 0) as the $d_{x^2-y^2}$
orbitals of Mn(1) and Mn(2), thus these 
orbitals and O $p_{x}$ and $p_{y}$ orbitals form $p$-$d$ $\sigma$ bond. As
the bond angle Mn(1)-O(1)-Mn2 is only 157$^o$, the strength of this covalent bond is
weak and consequently the AF superexchange interaction becomes weakened.\cite{goodenough69}
Despite the AF superexchange interaction, there is no exact cancellation of
the spins of Mn(1)$^{3+}$ and Mn(2)$^{2+}$, and the result is a Ferri state with a finite
magnetic moment of 0.85\,$\mu_B$.
\par
Transition-metal perovskite oxides which exhibit CO like La$_{1-x}$Sr$_x$MnO$_3$
have an octahedral crystal field, whereas YBaMn$_2$O$_5$ has a square-pyramidal arrangement 
around Mn. In this case the $d$ orbitals of Mn split
into low-lying $e_{g}$ orbitals and relatively higher-lying $t_{2g}$ orbitals.  
Fig.\,\ref{fig:orb-dos}, shows that $t_{2g}$ is closer to $E_{F}$ than $e_{g}$. 
The same feature is observed for the isostructural phase YBaCo$_2$O$_5$.\cite{kwon2000}
From crystal-structure considerations it is deduced that HOMO
(highest occupied molecular orbital) is located at the top of the bonding $\pi$
level of VB, arising from the Mn $t_{2g}$ orbitals with a band gap to the empty LUMO (lowest 
unoccupied molecular orbital) located at the bottom of the antibonding $\pi^{\star}$ level 
of CB.\cite{beales97} Among the $t_{2g}$ orbitals $d_{xz}$ and $d_{yz}$ of both Mn(1) and Mn(2)
are energetically degenerate as clearly seen from the character of the DOS curve.
(In YVO$_3$ also $d_{xz}$ and $d_{yz}$ are nearly degenerate\cite{sawada98}.)
So Mn(1) and Mn(2) exist in high-spin state with the $e_g$ orbitals filled up before the
$t_{2g}$ orbitals. The $d_{xz}$ and $d_{yz}$ orbitals contain one electron each for Mn(1)
and Mn(2). For Mn(1) a very small peak is seen, which could come from down-spin $d_{xy}$, 
whereas a finite sized peak is observed for the same orbital of Mn(2). It is indeed the 
occupancy of this orbital which determines the magnetic moment of Mn(1) and Mn(2). 
If one considers the Goodenough-Kanamori\cite{goodenough2} rules for magnetic interactions in manganese
oxides, the expected magnetic order should be A-type AF ({\em viz.} F ordering within the 
layers and AF ordering between the layers; arising from superexchange interactions
between occupied $d_{x^2-y^2}$ orbitals on Mn$^{2+}$ and empty $d_{x^2-y^2}$
on Mn$^{3+}$). However, owing to the large deviation of the Mn-O-Mn bond angle from 
180$^{o}$ along with CO the $d_{x^2-y^2}$ orbitals for both Mn species become
occupied. Hence, AF ordering is observed between Mn within the planes as
well as between the planes (see Fig.\,\ref{fig:order}). One Mn couples AF to its six 
neighboring Mn in a G-type AF arrangement in accordance with experimental 
findings.\cite{millange99} Hence, the theoretical calculations have provided the 
correct ground state with respect to the experiments.
\par
Magnetic susceptibility and magnetization measurements, have unequivocally shown that 
YBaMn$_2$O$_5$ is in a Ferri state at low temperature\cite{chapman96,millange99} with a
saturated moment between 0.5 and 0.95\,$\mu_{B}$ per YBaMn$_2$O$_5$ 
formula unit.\cite{chapman96,millange99} 
The theoretically calculated magnetic moments for Mn are 3.07 and 
3.93\,$\mu_B$, repectively, giving a net magnetic moment of 0.86\,$\mu_B$ per formula unit 
for the Ferri state of YBaMn$_2$O$_5$. Our theoretically calculated value is less than
the predicted (spin-only) value of 1.0\,$\mu_B$. From the DOS analyses, we noted
that there is strong hybridization between Mn $3d$ electrons and O $2p$ electrons.
A finite magnetic moment of 0.0064 and 0.0032\,$\mu_B$/atom are theoretically 
found to be present at O(1) and O(2), respectively. Hence, we conclude that the slight 
deviation
in the saturated magnetic moment from that predicted by an idealized ionic model can be
attributed to the strong hybridization between Mn $3d$ and O $2p$ electrons.
We have also performed calculations with the room temperature structural parameters,
confirming that the two manganese atoms possess different magnetic moments at
this temperature.

\subsection{Charge and orbital ordering}
\label{subsec:charge}
For the pseudo-cubic and layered perovskite manganese oxides, essentially  three 
parameters control the electron-correlation strength and the resultant structural, transport
and magnetic properties.\cite{kuwahara98} First, the hole-doping level 
(charge-carrier density
or the band-filling level of CB). In the case of perovskite
oxides the substitution of trivalent $RE$ by divalent $AE$ introduces holes in the Mn 
$3d$ orbitals. Second, the effective one-electron bandwidth ($W$) or equivalently the
$e_{g}$ electron-transfer interaction. The magnitude of $W$ is directly determined 
by the size of the atom at the $RE, AE$ site which makes the Mn-O-Mn bond angle 
deviate from
180$^o$ and thus hinders the electron-transfer interaction. The correlation between CO and
the size of $RE$ and $AE$ is studied by several workers and is well illustrated 
by a phase diagram in Ref.\,\onlinecite{rao98}. Third comes the dimensionality: the 
lowering of the electronic dimensionality causes a variety of essential changes in 
the electronic properties. The carrier-to-lattice coupling is so strong in manganites, that the
charge-localization tendency becomes very strong. In general the ground state of mixed-valent 
manganite perovskites is, therefore, either F and metallic or AF and CO. 
In all CO systems, the magnetic susceptibility drops rapidly at the
CO temperature ($T_{\rm{CO}}$). CO drastically influences the magnetic 
correlations in manganites.
Investigations on the CO state have established an intimate connection to lattice
distortion. It seems to be the lattice distortion associated with OO which
localizes the charge and thus initiates CO.\cite{rao98} 
The effect of the CO state on cooperative magnetic states is to produce insulating behavior.
A high magnetic field induces a melting-like phenomenon of the electron lattice of the CO 
phase giving rise to a huge negative magnetoresistance.\cite{tokura96} For these reasons, it is 
interesting to study CO in YBaMn$_2$O$_5$. 
\par
Charge localization, which is a prerequisite for CO is mutually exclusive with
an F state according to and the double-exchange mechanism. The double-exchange mechanism 
requires hopping of charge carriers from one Mn to an adjacent Mn via an intervening O.  
The CO state is expected to become stable when the repulsive
Coulomb interaction between carriers dominates over the kinetic energy of the carriers.
Hence, CO arises because the carriers are localized at specific long-range-ordered sites 
below  the CO temperature.
CO is expected to be favored for equal proportions of Mn$^{2+}$ and Mn$^{3+}$ as in 
the present case, and in YBaMn$_2$O$_5$ it is associated with the AF coupling between Mn 
in the $ab$ plane. CO does not occur in Pr$_{0.5}$Sr$_{0.5}$MnO$_3$\cite{kuwahara98} 
where A-type AF is the ground state, whereas CO is observed in 
Nd$_{0.5}$Sr$_{0.5}$MnO$_{3}$\cite{kuwahara98}
below 150\,K where CE-type AF is the ground state. CO depends on the $d$-electron bandwidth
and hence it is worth to consider this feature in some detail. On reduction of the 
Mn-O-Mn angle, the hopping between the Mn $3d$ and O $2p$ orbital decreases and hence 
the $e_{g}$ 
bandwidth decreases. Consequently the system stabilizes in a Ferri-CO-insulating state. 
Usually the CO-insulating state transforms to a metallic F state on the
application of a magnetic field. This may be the reason for the metallic
behavior of the F phase found in our calculation. CO in YBaMn$_2$O$_5$ is characterized by the
real-space ordering of Mn$^{2+}$ and Mn$^{3+}$ species.
Our calculations predict that a long-range CO of Mn$^{2+}$ and Mn$^{3+}$ with a 
rocksalt-type arrangement occurs at low temperatures. This can be viewed as chains 
of Mn$^{2+}$
and Mn$^{3+}$ running parallel to {\em b} and correspondingly alternating chains 
running along $a$ and $c$ ({\em viz.} a checker-board 
arrangement as seen from Fig.\,\ref{fig:order}).
\par
Furthermore, there exists orbital degrees of freedom for the e$_g$ electrons and
OO can lower the electronic energy through the JT mechanism.
Therefore, mixed-valent manganites can have OO in addition to CO.\cite{cheong98}
OO gives rise to the anisotropy in the electron-transfer interaction.
This favors or disfavors the double-exchange interaction and the (F or AF) superexchange 
interaction in an orbital-dependent manner and hence gives a complex spin-orbital-coupled
state. Therefore, it is also interesting to study OO in some detail.
Fig.\,\ref{fig:charge} shows the electron charge density of YBaMn$_2$O$_5$ in (001) and (110) 
planes. The electron-charge density is plotted in the energy range where the $3d$
orbitals reside, the shape of the $d_{x^2-y^2}$ and $d_{z^2}$ orbitals are well reproduced
in Fig.\,\ref{fig:charge}a and b, respectively.
In Fig.\,\ref{fig:charge}a, Mn(1), O(1) and Mn(2) are linked through covalent bonds which 
is easily seen as the finite electron density on the connecting lines between the atoms. 
This illuminates the path for the AF superexchange interaction between them. 
When the size of $RE$ and $AE$ becomes smaller the one-electron bandwidth (or
e$_g$ electron-transfer interaction) decreases in value.\cite{rao98} For 
Y$^{3+}$ with an ionic radius of 1.25\,\AA, [smaller than other $RE$s like La$^{3+}$ 
(1.36\,\AA)\cite{sawada98}] the Mn(1)-O(1)-Mn(2) angle is much less 
than 180$^o$ so the $e_g$ electron bandwidth is small compared with
the $t_{2g}$ electron bandwidth. Fig.\,\ref{fig:charge}a shows  
that despite the finite electron density between Mn and O, the $p$ orbitals
of O are not directed towards the lobes of the Mn $d_{x^2-y^2}$ orbitals. Hence, the
strength of the resulting $p$-$d$ covalent bond is decreased.
The orbital projected DOS in Fig.\,\ref{fig:orb-dos} shows that the  
$t_{2g}$ bandwidth is larger than the $e_g$ bandwidth owing to this hybridization effect.
\par
The transfer integral between the two neighboring Mn atoms in the crystal lattice
is determined by the overlap between the $3d$ orbitals with the $2p$ orbital of O atom.
Fig.\,\ref{fig:charge}b shows the electron density along (110)  of the unit cell. The 
$d_{z^2}$ orbital is ordered along $c$ and for both Mn(1) and Mn(2), this orbital 
hybridizes
with the O(1) $p_z$ orbital resulting in a $p$-$d$ $\pi$ bond. However, as the $d_{z^2}$ orbital 
forms a strong $\sigma$ bond with the $p_z$ orbital of O(2), the strength of this $\pi$ 
bond is weak.
The overlap between the $d_{x^2-y^2}$ and $p_z$ orbitals is zero because of their different
orientation in the $ab$ plane. Therefore, the electron in the $d_{x^2-y^2}$ orbital can not 
hop along $c$.\cite{goodenough69} In this manner the $e_g$ electrons get localized 
and cause CO and OO. 
Owing to the fact that the Mn(1)-O(1) bond length is 1.908\,{\AA} compared with 2.086\,{\AA}
for Mn(2)-O(1), more electronic charge is present on Mn$^{2+}$ than on Mn$^{3+}$. 
This is visible in the orbital decomposed DOS (Fig.\,\ref{fig:orb-dos}), where the $d_{xy}$
orbital of Mn$^{2+}$ has more states (electrons) than that of Mn$^{3+}$. In cubic 
perovskites, the electron 
transfer is almost prohibited along $c$ because of the orbital ordering of $d_{x^2-y^2}$, 
which is also the origin of the inter-plane AF coupling.\cite{toku_sci} 
\par
PND\cite{millange99} indicates that Mn$^{3+}$ has the occupied $d_{z^2}$ orbital extending 
along [001], whereas the unoccupied $d_{x^2-y^2}$
orbital extends along [110] and [1$\overline{1}$0]. A corresponding OO could be  
expected for Mn$^{2+}$ with both $d_{z^2}$ and $d_{x^2-y^2}$ orbitals occupied. However, our 
detailed electronic structure studies show that both $d_{z^2}$ and $d_{x^2-y^2}$ orbitals are 
partially occupied for Mn$^{2+}$ as well as Mn$^{3+}$ as shown in Fig.\,\ref{fig:orb-dos}. On the
other hand, according to our charge-density analysis (Fig.\,\ref{fig:charge}a and b) 
the $d_{z^2}$ orbital is ordered along [001]
and $d_{x^2-y^2}$ along [110] for Mn(1) and Mn(2) (Fig.\,\ref{fig:order}), which is 
consistent with experimental findings.
The Mn(1)-O-Mn(2) bond angle is much smaller than 180$^{o}$ which reduces
the effective $e_{g}$-$e_{g}$ hopping, the $e_{g}$ bands get localized and the
bandwidth reduced. This is the main reason for OO in YBaMn$_2$O$_5$.
Both $d_{z^2}$ and $d_{x^2-y^2}$ orbitals are aligned in the same orientation
within the layer as well as between the layers as shown in Fig.\,\ref{fig:order}. So,
this type of OO is named F type.

\subsection{Optical properties}
\label{subsec:optic}
Further insight into the electronic structure can be obtained from the calculated 
inter-band optical functions. It has been earlier found that the calculated optical properties for
SnI$_2$, NaNO$_2$ and Mn$X$ ($X$ = As, Sb, Bi)\cite{ravi1,ravi2,ravi3} are in excellent 
agreement with the experimental findings, and we have therefore used the same theory to predict the
optical properties of YBaMn$_2$O$_5$. Since, this material possesses unique Ferri
ordering and insulating behavior along with an uniaxial crystal structure it may find 
application in optical devices. Yet another reason for studying the optical properties is that, it has 
been experimentally established\cite{okimoto98} that the optical anisotropy of 
Pr$_{0.6}$Ca$_{0.4}$MnO$_3$ is drastically reduced above $T_{\rm{CO}}$. It is therefore expected that
the optical anisotropy will provide more insight about CO and OO in YBaMn$_2$O$_5$.
For YBaMn$_2$O$_5$ with its tetragonal crystal structure, the optical spectrum is conveniently
resolved into two principal directions $E\|a$ and $E\|c$ {\it viz.} with the electric field vector 
polarized along $a$ and $c$, respectively. In the top-most panel of Fig.\,\ref{fig:opt1},
the dispersive part of the diagonal elements of the dielectric tensor are given. The anisotropy
in the dielectric tensor is clearly seen in this illustration. 
\par
In the second panel of Fig.\,\ref{fig:opt1}, the polarized $\epsilon_{2}$ spectra are shown. 
The spectrum corresponding to $E\|a$ and $E\|c$ differ from one another up to
$ca.$ 10\,eV whereas less difference is noticable in the spectra above 10\,eV. Since there is 
an one-to-one correspondance between the inter-band transitions and band structures (discussed
in Sec.\,\ref{subsec:band}), we investigate the origin of the peaks in the $\epsilon_{2}$ spectrum
with the help of our calculated band structure. As YBaMn$_2$O$_5$ stabilizes in the Ferri state, 
VB has an unequal number of
bands in the up- and down-spin channels (Fig.\,\ref{fig:band}), $viz.$ 36 bands in 
the former and 38 in the latter.
The two extra bands of the down-spin channel closer to $E_F$ in VB play an important role in the 
transitions as discussed below. We name the top-most band of VB as no.\,38 and bottom-most band of
CB as no.\,39. The lowest-energy peak A results from inter-band transitions (no.\,35 to 41,
mostly O(1) $2p$ to Mn(2) $d_{z^2}$ and no.\,35 to 39, mostly Mn(1) $d_{z^2}$ to Mn(2) $4p$) and 
peak B
results from transitions (no.\,38 to 48 and 36 to 49, mostly Mn $3d$ to Mn $4p$). The peak C
originates from many transitions, including O $2p$ to Mn $3d$, Y $5s$ to Y $5p$ {\it etc.} 
Peaks D, E and
F are contributed by several transitions including O $2p$ to Mn $3d$, Y $5s$ to Y $5p$. Further,
a very small peak is present in the higher-energy region ($\sim$ 17\,eV) of $\epsilon_{2}$ 
which is
due to transitions from lower-lying occupied levels to higher-lying unoccupied levels. The 
accumulation of broad Y $4d$ and Ba $5d$ bands in the high-energy part of CB results in very
little structure in the higher-energy part of the optical spectra.
The optical gaps for $E\|a$
and $E\|c$ are approximately the same indicating that the
effective inter-site Coulomb correlation is the same for the in-plane and the out-of-plane 
orientation for the Ferri phase. This can be traced back to the G-type Ferri coupling 
in this material. 
\par
In order to understand the origin of the optical anisotropy in YBaMn$_2$O$_5$ we have 
also made optical property calculations for the F phase, which show that (in contrast to the
Ferri phase) the $\epsilon_{2}$ spectrum for $E\|a$ is shifted some 3.5\,eV to higher values
than $E\|c$. Further, the $\epsilon_{2}$ components of $E\|c$ for the F phase is much
smaller than that for the Ferri phase indicating that the large optical anisotropy
in YBaMn$_2$O$_5$ is originating from the G-type Ferri ordering. 
\par
To emphasize the above
finding, we have also plotted the spin-projected $\epsilon_{2}$ spectra along the $a$ 
[$\epsilon_{2}^{a}(\omega)$] and $c$ [$\epsilon_{2}^{c}(\omega)$]
directions (third and fourth panels of Fig.\,\ref{fig:opt1}, respectively). Although the optical 
gap is approximately
same for $E\|a$ and $E\|c$ in the $\epsilon_{2}$ spectrum, there is a finite 
difference in the optical gaps related to up- and down-spin electrons in the 
$\epsilon_{2}^{a}(\omega)$ and $\epsilon_{2}^{c}(\omega)$ spectra. The optical gap for the 
down-spin case is smaller than that for
up-spin case owing to the presence of the two narrow bands very close to $E_F$ in the down-spin 
channel of VB.
There is a large difference between the spectra for up- and down-spins up to $ca.$ 7\,eV.
The $\epsilon_{2}^{a}(\omega)$
spectrum resulting from the up-spin states has somewhat more dispersed peaks than that from
down-spin states. The $\epsilon_{2}^{a}(\omega)$           
spectrum resulting from the down-spin states has four well-defined peaks;
two prominent peaks in the region 1.75 to 2\,eV, and two additional peaks at $ca.$ 2.25 and
3\,eV. The magnitude of the down-spin peaks are higher than those of the up-spin peaks in 
the $\epsilon_{2}^{a}(\omega)$ spectrum.
The $\epsilon_{2}^{c}(\omega)$ spectrum originating from up- and down-spin states have appreciable
differences up to $ca.$ 6\,eV. The down-spin part has two well-defined peaks at $ca.$ 
1.75 and 2.75\,eV. The up-spin part has dispersed peaks of lower magnitude 
than the down-spin part, the magnitude of the up-spin peaks in 
$\epsilon_{2}^{c}(\omega)$ being generally higher than in $\epsilon_{2}^{a}(\omega)$.  
The optical anisotropy is noticable in the direction- as well as spin-resolved $\epsilon_{2}$
spectra. Hence, it is verified that the optical anisotropy originates both from crystal field
effects as well as from the Ferri ordering. As reflectivity, absorption coefficient and 
refractive 
index are often subjected to experimental studies, we have calculated these quantities and 
reproduced them in Fig.\,\ref{fig:opt2}. We now advertice for 
experimental optical studies on YBaMn$_2$O$_5$. 
 
\subsection{XANES studies}
\label{subsec:xane}
X-ray absorption spectroscopy (XAS) has developed into a powerful tool for the elucidation
of the geometric and electronic structure of amorphous and crystalline solids.\cite{behrens1}
X-ray absorption occurs by the excitation of core electrons, which makes this 
technique element specific. Although the X-ray absorption near edge structure (XANES) 
only provides direct information about the unoccupied electronic states of a material, 
it gives indirect information about the valence of a given atom in its particular environment 
and about occupied electronic states. This is because the unoccupied states are affected 
by the occupied states through interaction with the neighbors. 
\par
\begin{figure}
\caption{Calculated $K$-edge spectra for Mn(1) and Mn(2) as well as
O(1) and O(2) of YBaMn$_2$O$_5$.}
\label{fig:xane}
\end{figure}
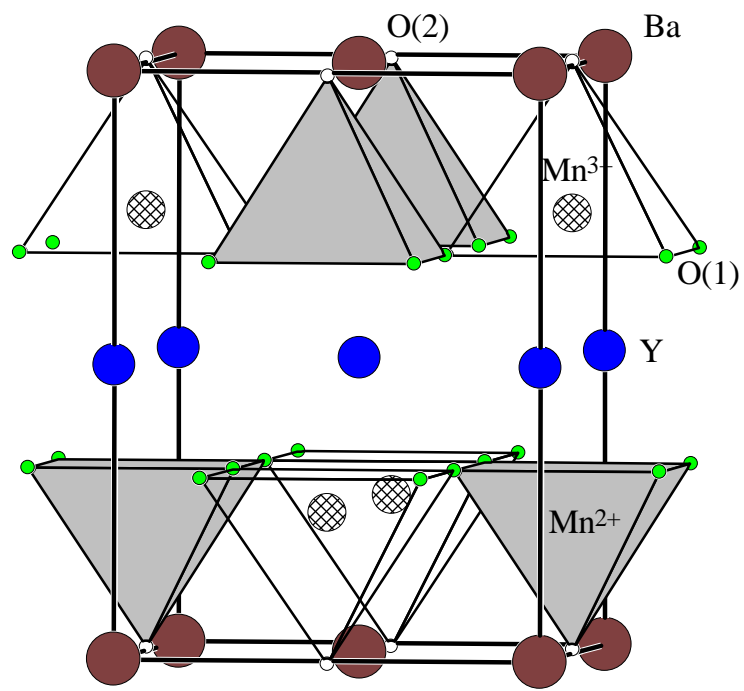
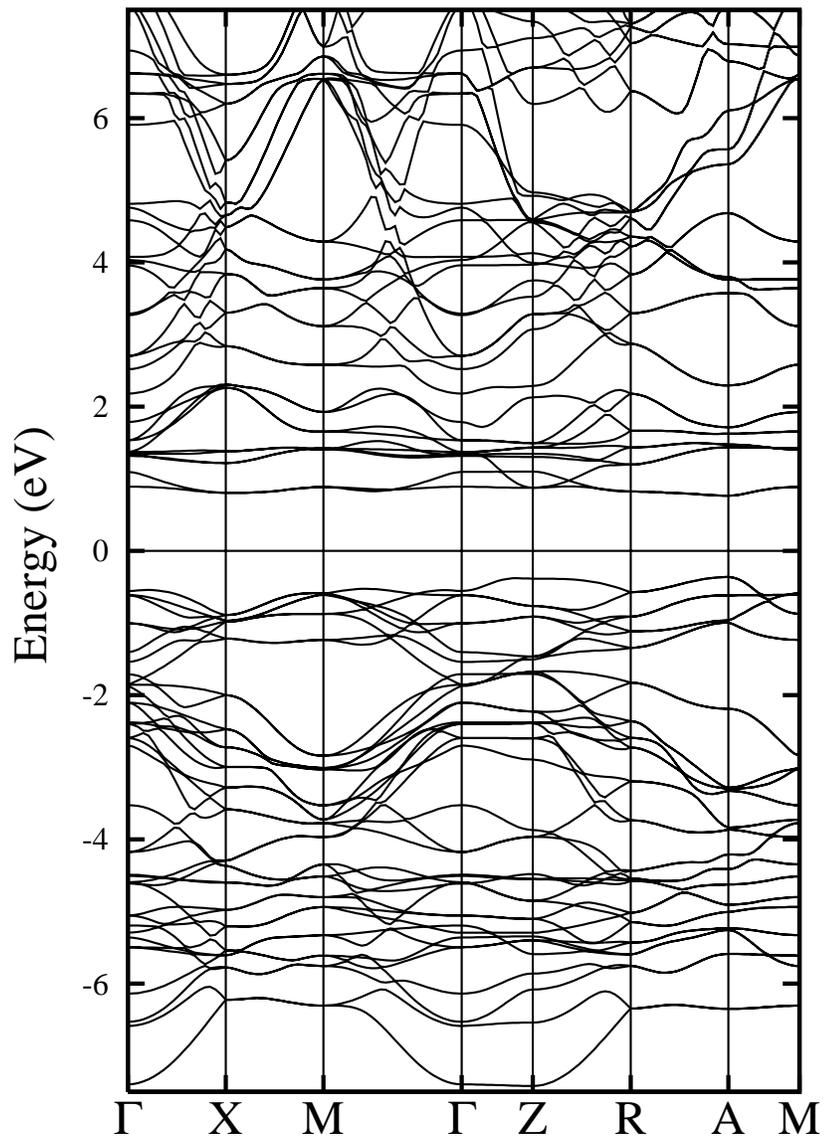
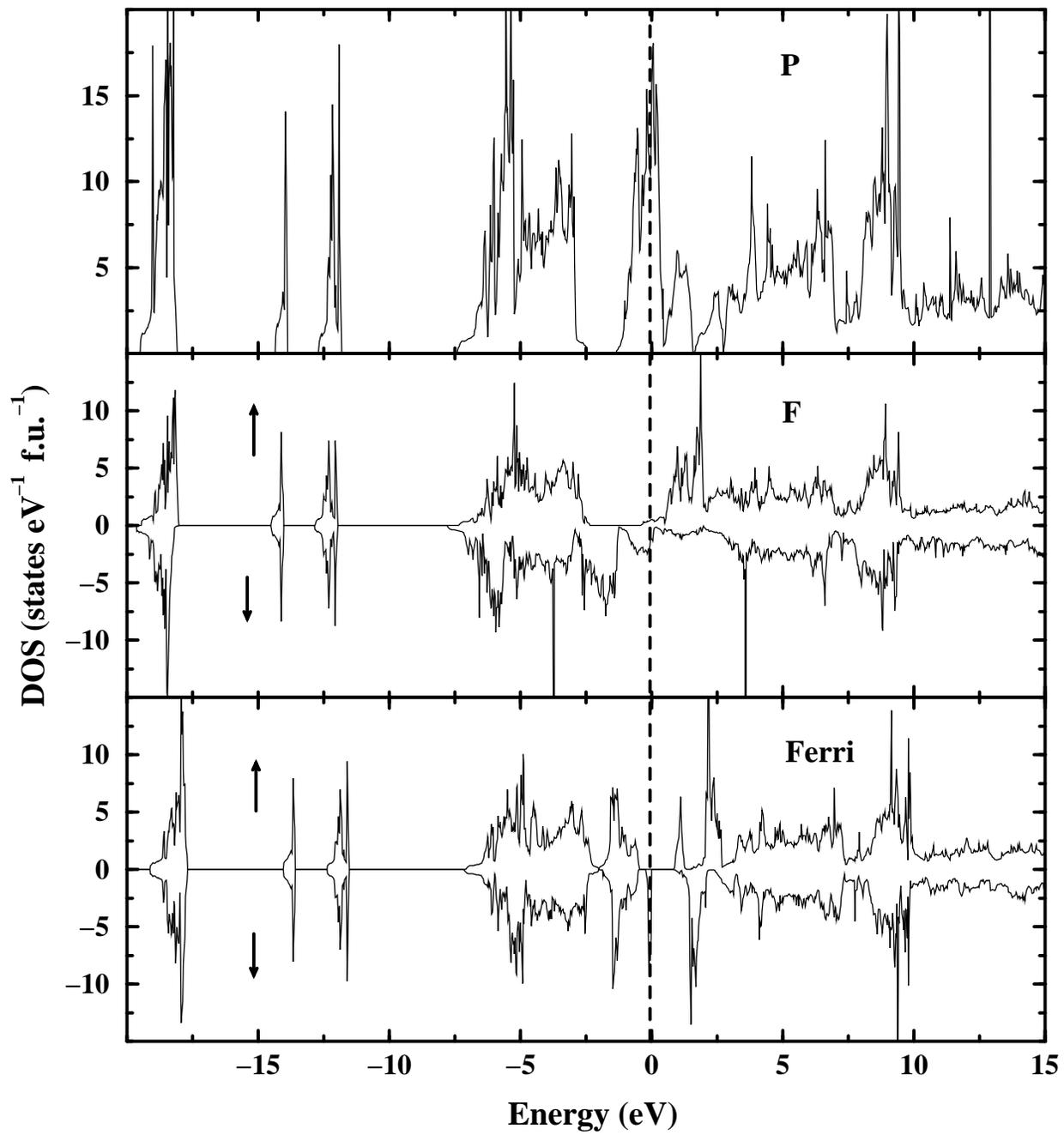
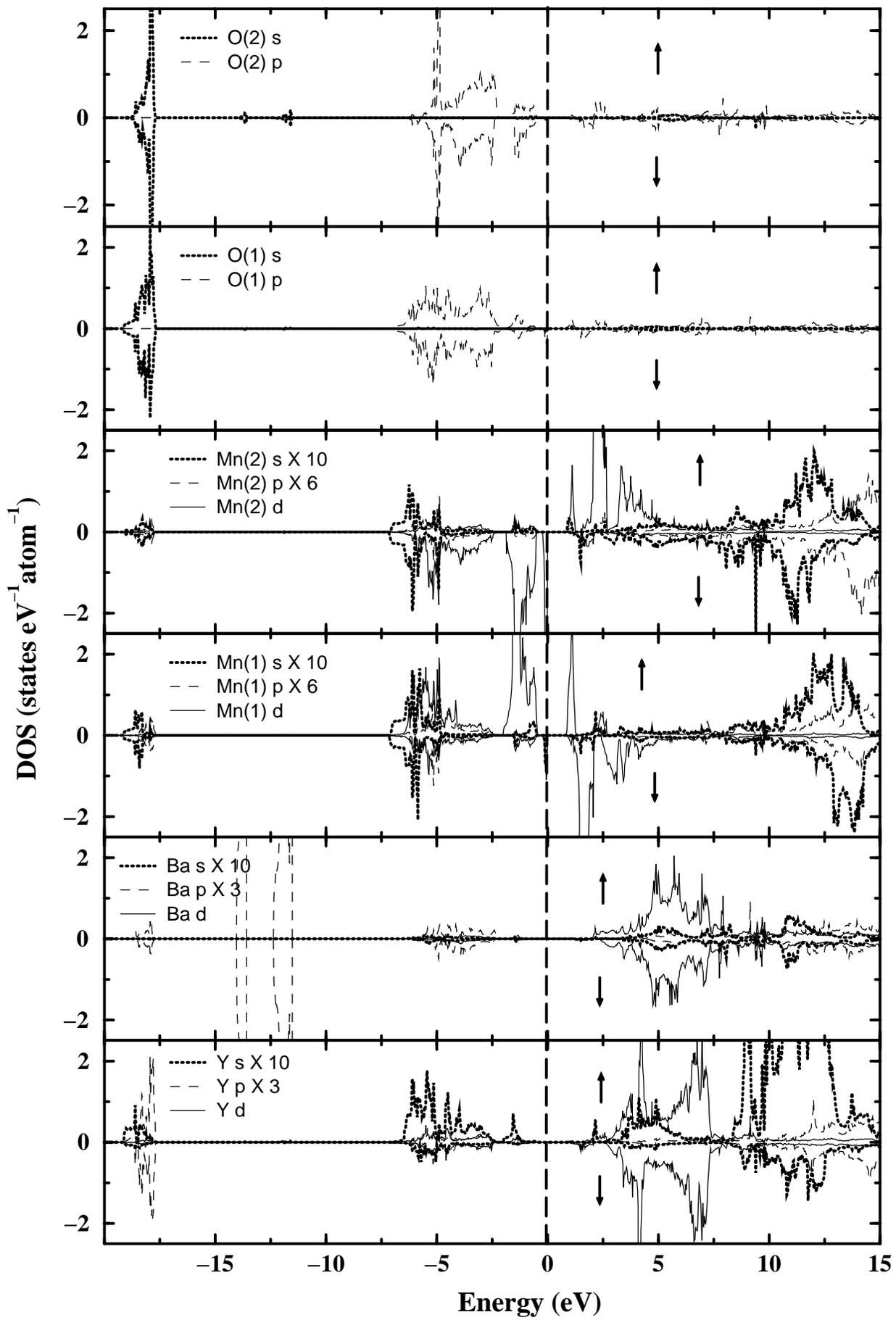
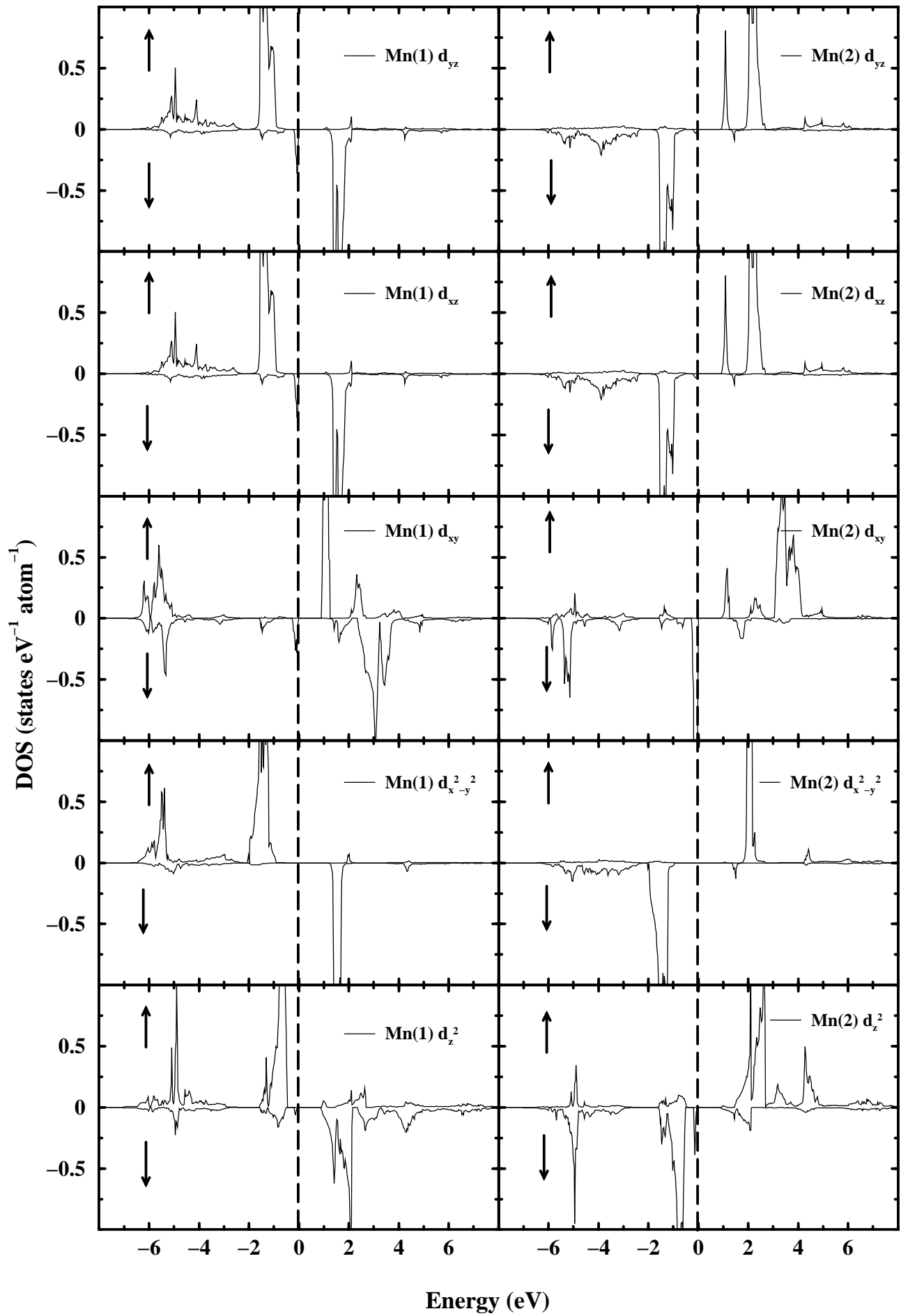
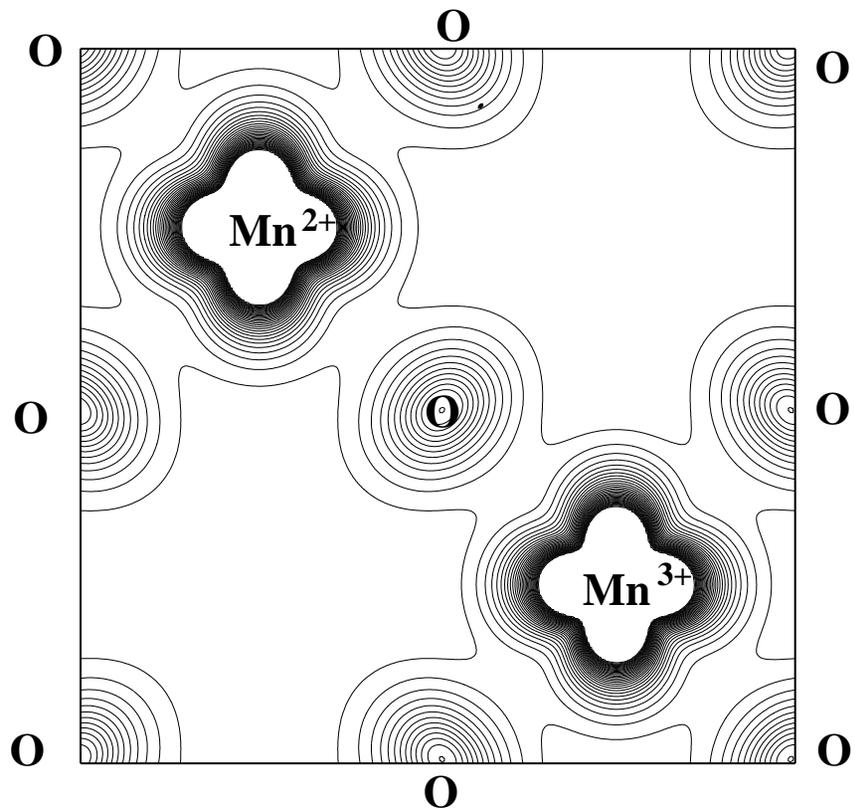
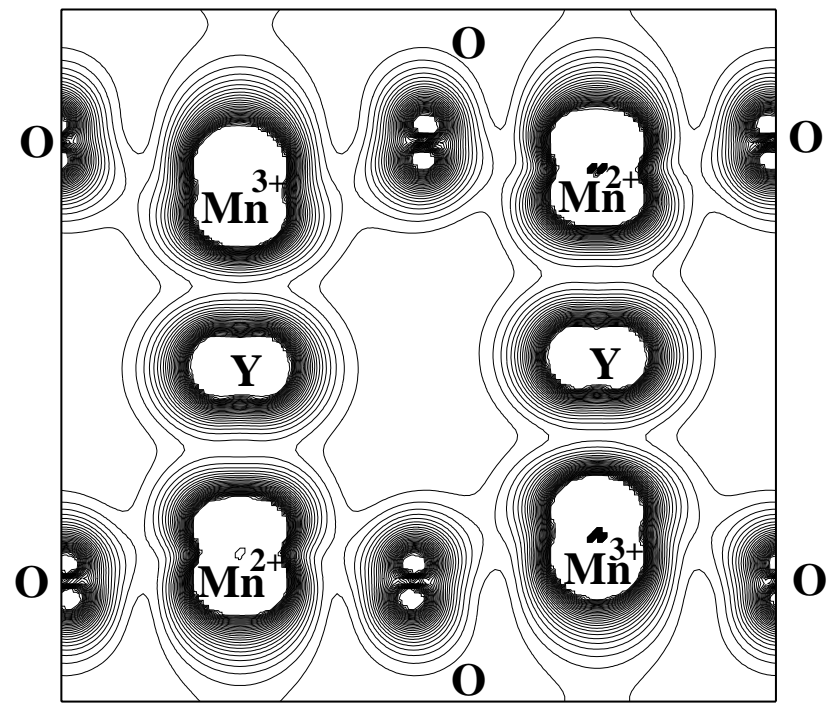
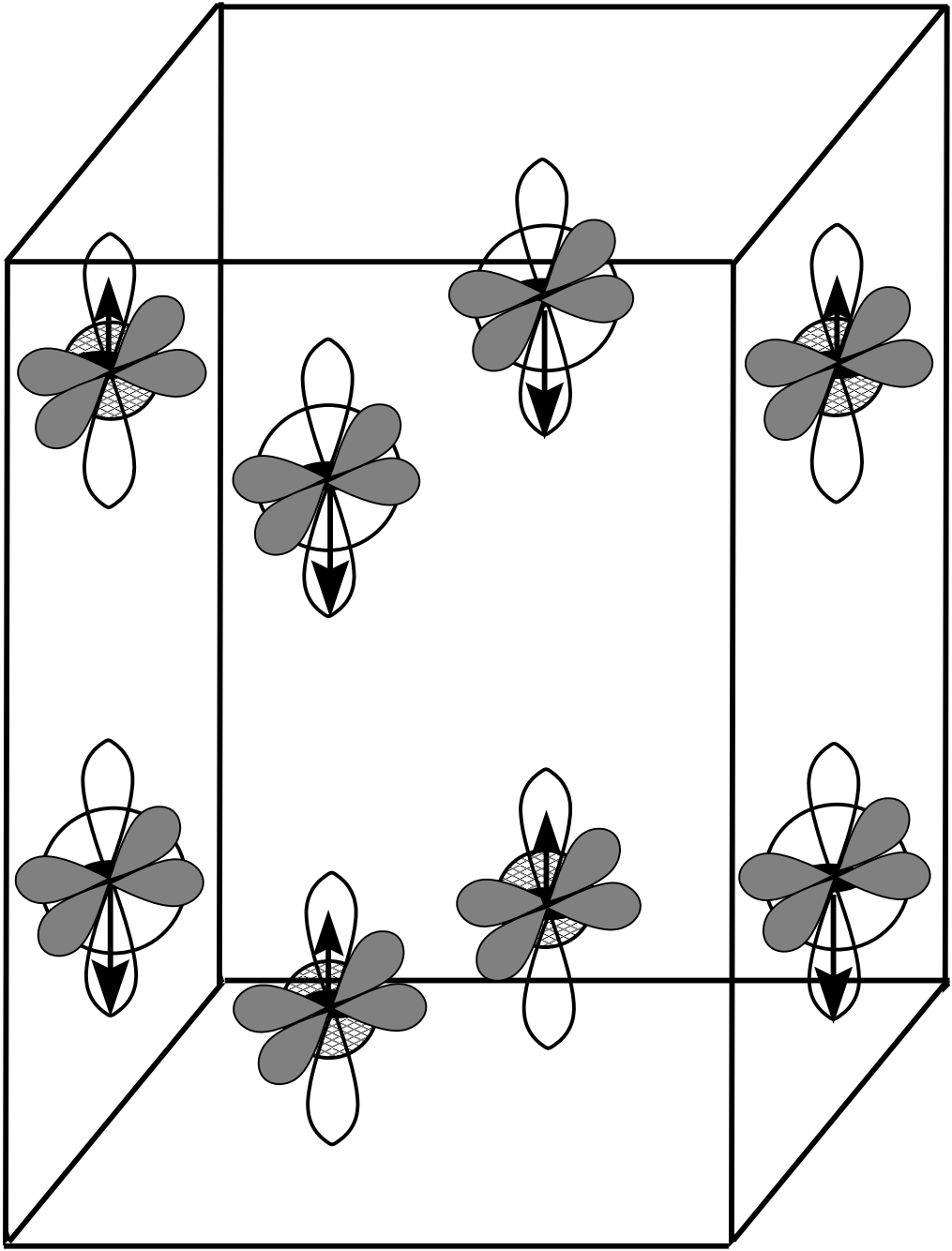
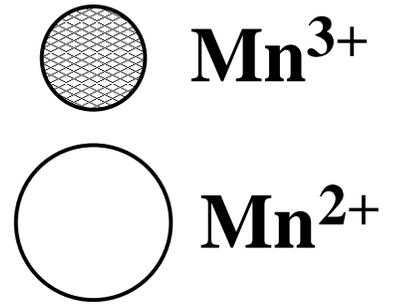
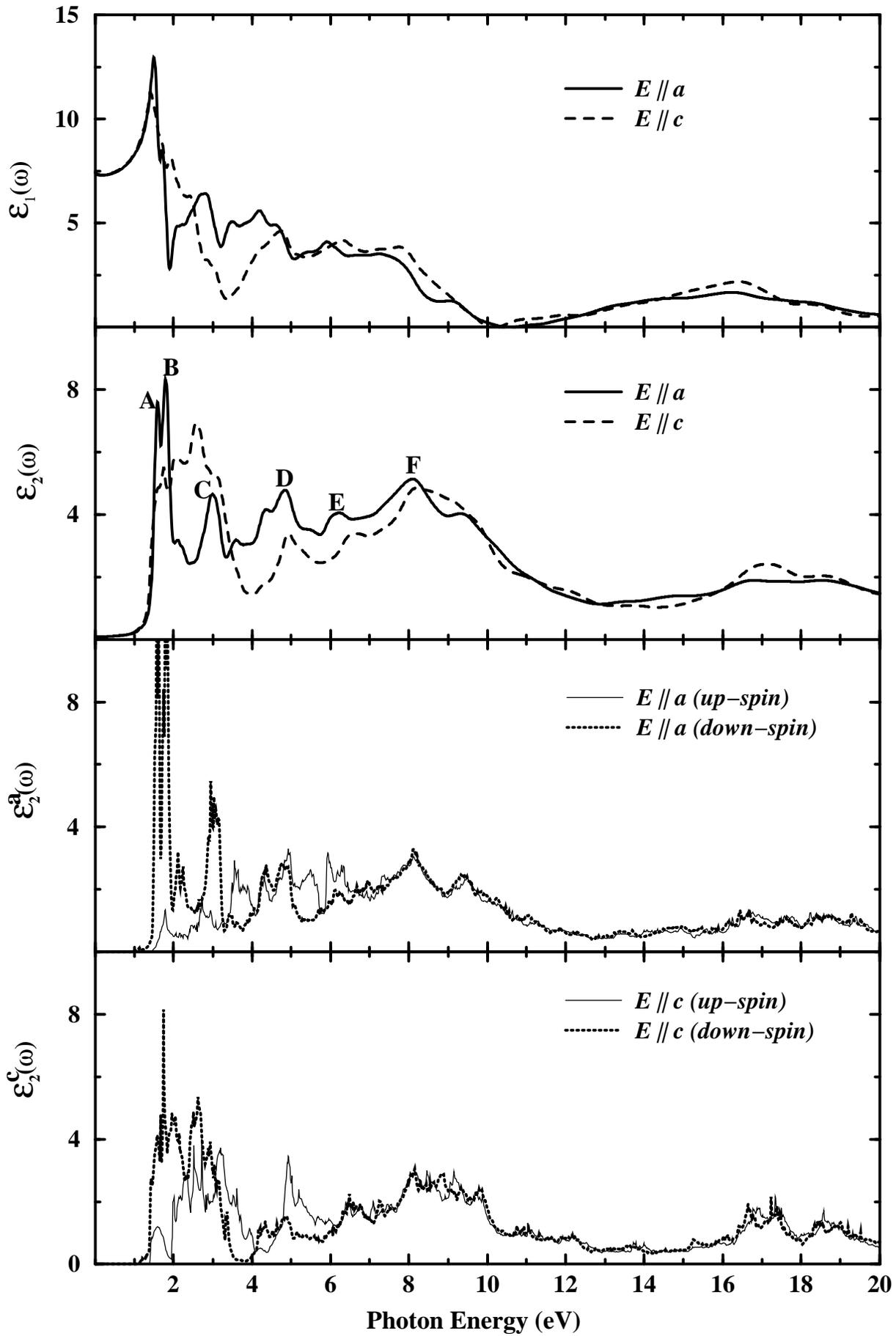
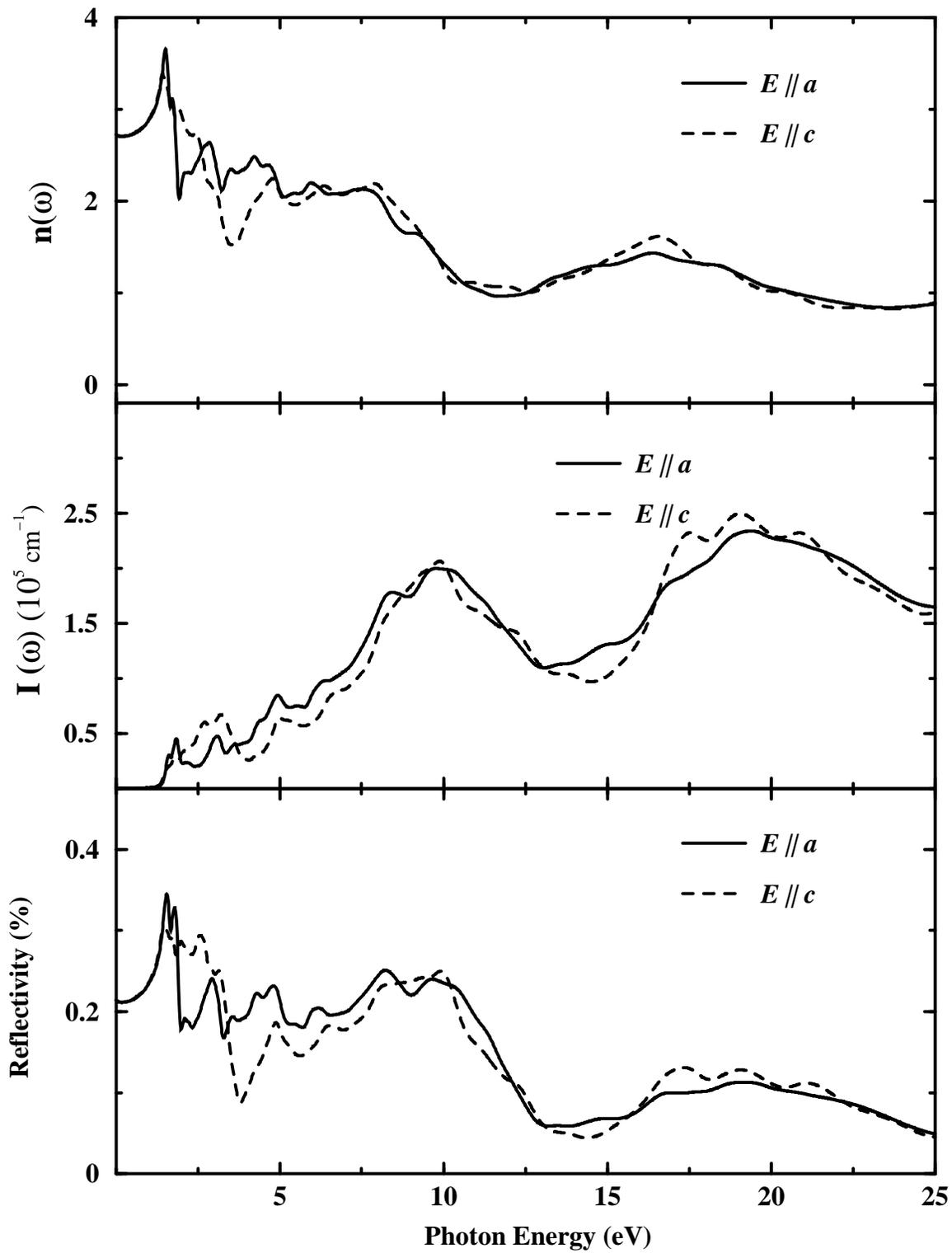
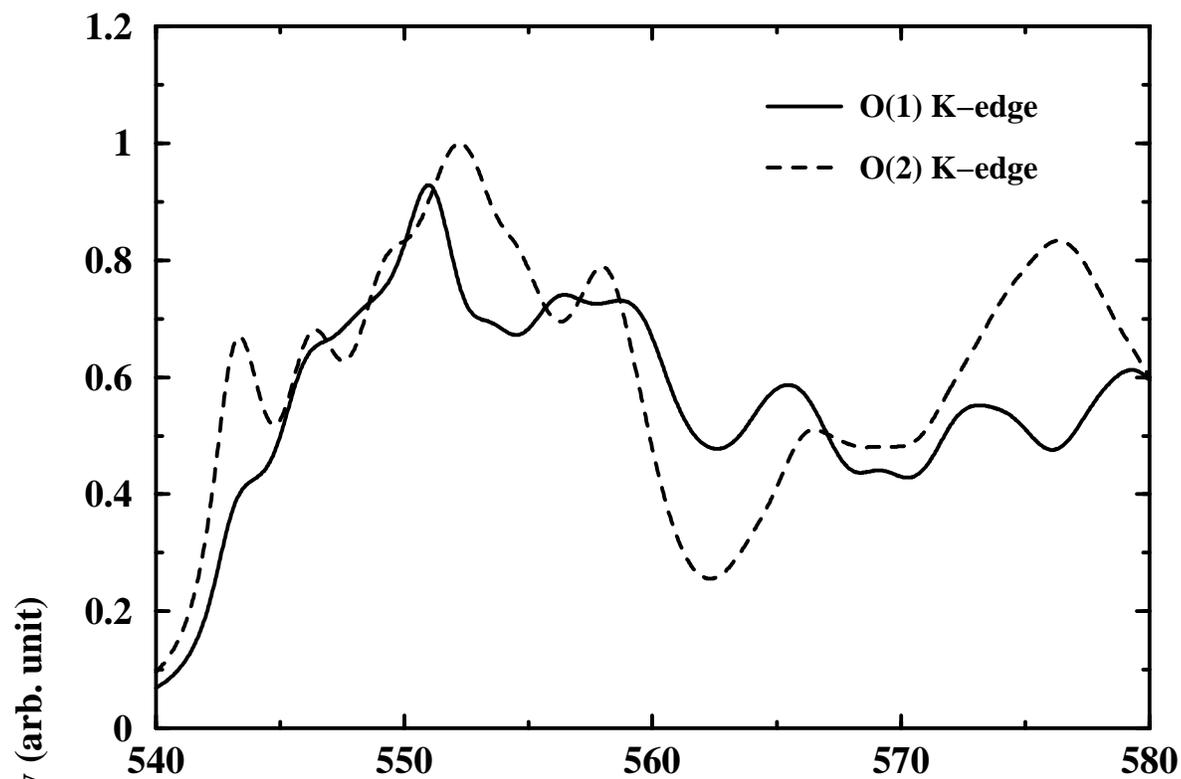
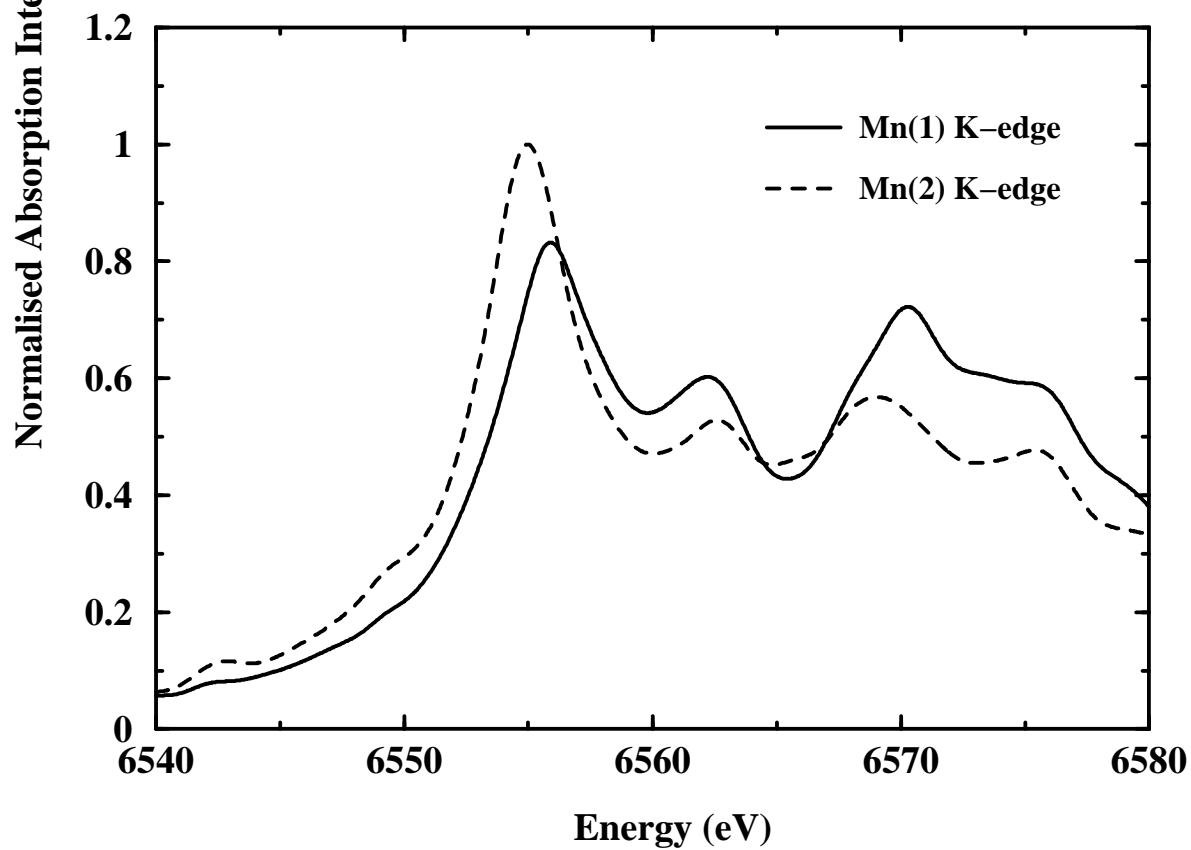

The oxygen atoms are in two different chemical environment in YBaMn$_2$O$_5$ as 
clearly seen in PDOS in Fig.\,\ref{fig:sitedos}. The calculated 
$K$-edge spectra for O(1) and O(2) shown in Fig.\,\ref{fig:xane} involve transition from the $1s$ 
core state to the unoccupied $p$ state. In this context the Mn $K$-edge mainly probes the
unoccupied Mn $4p$ states. It is generally accepted that O $K$-edge spectra is very sensitive 
to the local structure of transition-metal oxides as documented for 
Fe$_2$O$_3$\cite{paterson90} and TiO$_2$\cite{lusvardi98}.
The $K$-edge spectra for O(1) and O(2) (Fig.\,\ref{fig:xane}) show appreciable differences
throughout the whole energy range. In particular there are two
peaks appearing in the O(2) spectrum between 540 and 550\,eV that are absent for O(1).
Within 3\,\AA, 2 Mn, 2 Y and 6 O surround O(1) whereas, 2 Mn, 4 Ba and 
5 O surround O(2). It is this different chemical environment which causes the
differences in the $K$-edge XANES spectra of O(1) and O(2).
\par
YBaMn$_2$O$_5$ contains Mn in the valence states Mn$^{3+}$ and Mn$^{2+}$ which as 
discussed in Sec.\,\ref{subsec:charge}, experience CO. An experimental technique to visualize CO 
is not available. In order to visualize the presence of different oxidation states for Mn, 
we have theoretically calculated the XANES $K$-edge spectra for these and presented them
in Fig.\,\ref{fig:xane}. Both Mn atoms are seen to have four peaks within the energy range 
considered, reflecting that both are surrounded by five O within 2.08\,\AA. However, owing to 
the different valence states there are intensity differences as well as energy shifts (some 
1\,eV) of these peaks. For example, the lower-energy peak has large intensity in the 
Mn(2) $K$-edge 
spectrum compared with that for Mn(1). On the contrary, the three higher-energy peaks in the 
Mn(2) $K$-edge 
spectrum are less intense than in the Mn(1) $K$-edge spectrum. When experimental XANES spectra
become available for YBaMn$_2$O$_5$ the above features should be able to confirm the two 
different valence states for Mn.

\subsection{Hyperfine parameters}
\label{subsec:hyp}
The calculation of hyperfine parameters is useful to characterize different
atomic sites in a given material. Many experimental techniques such as
M\"{o}ssbauer spectroscopy, nuclear magnetic and nuclear quadrupole
resonance and perturbed angular correlation measurements 
are used to measure hyperfine parameters. Hyperfine parameters 
describe the interaction of a nucleus with electric and 
magnetic fields created by its chemical environment.  
The resulting splitting of nuclear energy levels is determined by the
product of a nuclear and an extra-nuclear quantity. In the case of quadrupole
interaction, it is the nuclear quadrupole moment that
interacts with the electric-field gradient (EFG) produced by the charges outside
the nucleus\cite{petrelli98}. EFG is a ground state property of a material 
which depends sensitively on the asymmetry of the electronic charges.
The direct relation of EFG and the asphericity of the electron density in
the vicinity of the probe nucleus enables one to estimate the quadrupole
splitting and the degree of covalency or ionicity of the chemical bonds provided
the nuclear quadrupole moment is known. Quantities describing hyperfine
interactions ($e.g,$ EFG and isomer shift) are widely studied nowadays both
experimentally and theoretically. 
\par
Blaha {\em et al.}\cite{efg85} have showed
that the linear augmented plane wave (LAPW) method is able to predict 
EFGs in solids with high precision. The charge 
distribution of complex materials such as YBa$_2$Cu$_3$O$_7$, YBa$_2$Cu$_3$O$_{6.5}$
and YBa$_2$Cu$_3$O$_6$ have been studied theoretically by Schwarz {\em et al.}\cite{efg2} 
by this approach.
In this study, we have attempted to establish the different valence states of Mn
in YBaMn$_2$O$_5$ with the help of EFG and the hyperfine field calculated
using FPLAPW method as embodied in the WIEN97 code.\cite{blaha97}
\par
The total hyperfine field (HFF) can be decomposed in three terms: 
a dominant Fermi contact term, a dipolar term and an orbital contribution.
We limit our consideration to the contact term, which in the non-relativistic limit is
derived from the spin densities at the nuclear site:

\begin{eqnarray}
H_{c} = \frac{8}{3}\pi\mu_{B}^{2}[\rho_{\uparrow}(0)-\rho_{\downarrow}(0)]
\end{eqnarray}
\par
EFG is defined as the second derivative of the electrostatic potential at the
nucleus, written as a traceless tensor. This 
tensor can be obtained from an integral over the non-spherical charge
density $\rho(r)$. For instance the principal component $V_{zz}$ is given
by
\begin{eqnarray}
V_{zz} = \int d^3\!r \rho({\bf r}) \frac{2P_{2}(cos\theta)}{r^{3}}
\end{eqnarray}
where $P_2$ is the second-order Legendre polynomial. A more detailed
description of the calculation of EFG can be found elsewhere.\cite{efg3}
\par
The calculated EFG and HFF at the atomic sites in
YBaMn$_2$O$_5$ are given in Table\,\ref{table:table2} which confirm that
there is a finite difference in the value of both EFG and HFF between the two
Mn atoms. So we can conclude that their charge distribution is quite different. The 
higher value of EFG and HFF in Mn$^{2+}$ than in Mn$^{2+}$ is justified 
because, more charge is found on the former. This can be seen from the orbital projected DOS 
as well as from the magnetic moments possessed by the two ions. HFF for Mn$^{3+}$
in LaMnO$_3$ is found to be $-$198\,kG\cite{ravilamn} which is 
quite close to $-$179\,kG found for Mn$^{3+}$ in our case. Consequently we substantiate that 
Mn(1) corresponds to Mn$^{3+}$. The two oxygens ions also differ mutually in their values for
EFG and HFF (Table\,\ref{table:table2}) suggesting that the strength of the covalent 
bond formed by them with Mn(1) and Mn(2) is different.

\section{SUMMARY}
\label{sec:sum}
Like hole-doped $RE$MnO$_3$-based CMR materials YBaMn$_2$O$_5$ also carries mixed-valence
states of manganese, ferrimagnetic ordering, charge ordering and apparently undergoes 
a combined insulator-to-metal and Ferrimagnetic-to-Ferromagnetic transition. Hence 
YBaMn$_2$O$_5$ may be a potential CMR material which deserves more attention.

We have made a detailed investigation on the electronic properties of YBaMn$_2$O$_5$ 
using the full potential LMTO method as well as the full potential LAPW method and conclude the
following. 

1. The G-type ferrimagnetic insulating state is found to be the ground state in accordance
with experimental findings.\\

2. The existence of the two different types of Mn atoms is visualized by differences in
site- and orbital- projected DOS curves. In order to further emphasize the different valence 
states of Mn, we have calculated $K$-edge XANE spectra. For Mn as well as O the existence of two 
types of valence induced atomic species is established with the help of $K$-edge spectra.\\

3. The occurrence of checker-board-type charge ordering and F-type orbital ordering 
is seen from the charge-density plots. The small size of Y$^{3+}$ makes the 
Mn-O-Mn bond angle deviate from 180$^o$, which in turn imposes a reduction in the e$_g$ 
bandwidth. The charge- and orbital-ordering features are believed to result from this 
perturbation of the $e_{g}$ orbitals.\\

4. As YBaMn$_2$O$_5$ is an ferrimagnetic insulator, it is useful to probe its optical properties
for potential applications. We have analyzed the inter-band contributions to the optical 
properties with the help of the calculated electronic-band-structure features. We found large
anisotropies in the optical spectra originating from ferrimagnetic ordering and the crystal
field splitting. No 
experimental optical study of YBaMn$_2$O$_5$ is hitherto available. \\

5. Hyperfine parameters such as hyperfine field and electric-field gradients have also been
calculated showing very large differences in the computed values for the 
crystallographically different manganese and oxygen atoms. This substantiates that Mn 
exist in two different valence states in YBaMn$_2$O$_5$.\\
\acknowledgements
This work has received support from The Research Council of Norway (Programme for Supercomputing)
through a grant of 
computing time. RV wishes to thank Dr. John Wills and Prof. Karlheinz Schwarz for supplying their
computer codes used in this study and Prof. Olle Eriksson for useful communications. 
RV kindly acknowledges Dr. Leif Veseth and P. Vajeeston for the useful discussions.

\newpage
\begin{table}
\caption{ Total energy in different 
magnetic configurations}
\begin{tabular}{ccccc}
Type & Paramagnetic & Ferromagnetic & Antiferromagnetic\\
\hline
Total energy (Ry/f.u.) & $-$28438.0466 & $-$28438.2754 & $-$28438.3125\\
$\Delta$E (meV/f.u.)     & 3618.6 & 505.6 & 0.0\\
Electronic state & Metal & Metal & Insulator\\
\hline
\hline
          &     &Calculated Magnetic Moment ($\mu_{B}$)\\
\hline
\hline
Magn. mom.  & Ferromagnetic & Antiferromagnetic & Experimental\cite{millange99}\\
\hline
Mn$^{3+}$ & 3.45 & 3.07 & 2.91\\
Mn$^{2+}$  & 3.99 & 3.93 & 3.91\\
Saturated    & 7.44 & 0.86 & 0.95\\
\end{tabular}
\label{table:table1}
\end{table}
\newpage
\begin{table}
\caption {Calculated (FLAPW method) principal component of the electric field 
gradient (EFG) $V_{zz}$ in units of 10$^{21}$ V/m$^2$ and the Fermi contact hyperfine
field (HFF) in kG at the atomic sites in ferro- and ferrimagnetic configurations.}
\begin{tabular}{||l|l|l|l|l||}
 & \multicolumn{2}{l|}{Ferromagnetic} &
 \multicolumn{2}{l||}{Ferrimagnetic} \\
\cline{1-5}
Atom  &{EFG}  &  {HFF}     &
  {EFG}  &  {HFF} \\
\hline
Y & 0.17 & 45.90 & 0.48 & 21.01 \\
Ba & 10.50 &$-$19.24 & 10.08 & 38.85  \\
Mn(1)  & 2.41 &$-$73.36 &  0.09 &$-$179.54 \\
Mn(2)  &$-$0.98 & $-$210.98 & 1.01 & 318.79 \\
O(1)  & 5.36 &  64.42 & 4.95 &$-$42.59 \\
O(2)  & 3.86 & 97.06 &  5.06 & 21.96 \\
\end{tabular}
\label{table:table2}
\end{table}
\begin{figure}
\caption{The YBaMn$_2$O$_5$ crystal structure. The
Y and Ba ions form layers along $c$. The square
pyramids corresponding to Mn$^{3+}$ and Mn$^{2+}$ are 
shown by open and shaded polyhedra, respectively. The O atoms
at the bases and apices of the square pyramids are denoted O(1) and O(2),
respectively.}
\label{fig:struc}
\end{figure}
\begin{figure}
\caption{The electronic band structure of YBaMn$_2$O$_5$ in 
the antiferromagnetic state; (a) up-spin bands and (b) 
down-spin bands. The line at 0\,eV
refers to the Fermi energy level.}
\label{fig:band}
\end{figure}
\begin{figure}
\caption{Total density of states (DOS) for YBaMn$_2$O$_5$ in
para-, ferro- and ferrimagnetic states.}
\label{fig:totdos}
\end{figure}
\begin{figure}
\caption{Site and angular momentum decomposed DOS of YBaMn$_2$O$_5$
in the ferrimagnetic state, obtained by the full-potential LMTO
method.}
\label{fig:sitedos}
\end{figure}
\begin{figure}
\caption{Orbital ($d$) decomposed DOS for Mn(1) and Mn(2)  
in the ferrimagnetic state obtained by the full-potential FLAPW
method.}
\label{fig:orb-dos}
\end{figure}
\begin{figure}
\caption{Valence-charge distribution of YBaMn$_2$O$_5$ in (a) (001)
and (b) (110) plane. 75 contours are drawn between 0.01 and
0.1 electrons/a.u.$^3$}
\label{fig:charge}
\end{figure}
\begin{figure}
\caption{Schematic diagram showing the checker-board-type
charge order (CO), F-type orbital order (OO) and G-type ferrimagnetic spin
order (SO) of Mn in YBaMn$_2$O$_5$.}
\label{fig:order}
\end{figure}
\begin{figure}
\caption{Imaginary parts and real parts of the optical dielectric tensor of YBaMn$_2$O$_5$
are given in the first and second panels, respectively (note that the spectra are 
broadened). The spin-projected imaginary parts of dielectric tensor along $a$
and $c$ directions are given in third and fourth panels, respectively.}
\label{fig:opt1}
\end{figure}
\begin{figure}
\caption{Calculated reflectivity spectra, absorption coefficient
[($I(\omega)$ in 10$^{5}$ cm$^{-1}$)], and refractive index 
[$n(\omega)$] along $a$ and $c$ for YBaMn$_2$O$_5$.}
\label{fig:opt2}
\end{figure}
\end{document}